\documentclass[12pt]{article}
\usepackage{epsfig, amsmath, amsfonts,  amsthm, natbib}
\usepackage{longtable}
\usepackage{adjustbox,lipsum}
\usepackage{authblk}
\usepackage{setspace}
\usepackage{tikz}
\usepackage{mathrsfs}
\usepackage{graphicx,psfrag,epsf}

\def\bm#1{\boldsymbol{#1}{}}

\newcommand{\bmu}{\bm{\mu}}
\newcommand{\bSigma}{\bm{\Sigma}}
\newcommand{\bOmega}{\bm{\Omega}}

\theoremstyle{plain}
\newtheorem{thm}{Theorem}[section]
\newtheorem{lem}{Lemma}
\newtheorem{proposition}[thm]{Proposition}

\newtheorem{rmk}{Remark}[section]

\newcommand{\bfx}{{\bf x}}
\newcommand{\bx}{{\bf x}}

\title{ Quantile Graphical Models:  Bayesian Approaches}
\author{Nilabja Guha$^{+}$, Veera Baladandayuthapani ${^{++}}$,Bani K. Mallick$^{*}$}
\affil{$^+$ Department of Mathematical Sciences, University of Massachusetts Lowell, Lowell, MA 01854, USA.
}
\affil{$^*$ Department of Statistics, Texas A \& M University, College Station, TX 77843, USA.
}
\affil{$^{++}$ Department of Biostatistics,  University of Michigan, Ann Arbor, MI 48103, USA.
}

\begin{document}

\maketitle 
\begin{abstract}
Graphical models are ubiquitous tools to describe the interdependence between variables measured simultaneously such as large-scale gene or protein expression data.   Gaussian graphical models (GGMs) are well-established tools for probabilistic exploration of dependence structures using precision matrices and they are generated under a multivariate normal joint distribution. However, they suffer from several shortcomings since they are based on Gaussian distribution assumptions. In this article, we propose a   Bayesian quantile based approach for sparse estimation of graphs. We demonstrate that the resulting graph estimation is robust  to outliers and applicable under general distributional assumptions. Furthermore, we develop efficient variational Bayes approximations to scale the methods for large data sets. Our methods are applied to a novel cancer proteomics data dataset where-in multiple proteomic antibodies are simultaneously assessed on tumor samples using reverse-phase protein arrays (RPPA) technology. 
\end{abstract}

\section{Introduction}
Probabilistic graphical models are the basic tools to represent  dependence structures among multiple variables. They provide a simple way to visualize the structure of a probabilistic model as well as provide insights into the properties of the model, including conditional independence structures.  A graph comprises with vertices (nodes) connected by edges (links or arcs). In a probabilistic graphical model, each vertex represents a random variable (single or vector) and the edges express probabilistic relationship between these variables. The graph defines the way the joint distribution over all the random variables can be decomposed into a product of factors contacting subset of the variables. There are two types of probabilistic graphical models:  (1)  Undirected graphical models where the edges do not carry the directional information (Sch\"{a}fer and Strimmer, 2005; Dobra et al., 2004; Yuan and Lin, 2007); (2) The other major class of graphical models is the directed graphical models (DAG)  or Bayesian networks where the edges of the graphs have a particular directionality which expresses causal relationships between random variables ({Friedman, 2004}; Segal et al., 2003; Mallick et al. 2009). In this paper, we  focus on the undirected graphical models.     

One popular tool of undirected graphical models is Gaussian Graphical Models (GGM) which  assume that the stochastic variables  follow a multivariate normal distribution with a particular structure of the inverse of the covariance matrix, called the precision or the concentration matrix. This precision matrix of the multivariate normal distribution has the interpretation of the conditional dependence. Compared with the marginal dependence, this conditional dependence can capture the direct link between two variables when all other variables are conditioned on. Furthermore, it is usually assumed that one of the variables can be predicted by those of a small subset of other variables. This assumption leads to sparsity (many zeros) in the precision matrix and reduces the problem to well known covariance selection problems ({Dempster, 1972;  Wong et al.,2003}).   Sparse estimation of precision matrix, thus plays a center role in Gaussian graphical model estimation problem (Friedman et al., 2008).

There has been an intense development of Bayesian graphical model literature over the past decades but mainly in a Gaussian graphical model setup. In a Bayesian setup, this joint modeling is done by  hierarchically specifying  priors on inverse covariance matrix (or precision matrix) using global priors on the space of positive-definite matrices. This prior specification is done through   inverse Wishart priors or hyper-inverse Wishart priors (Lauritzen,1996). Wishart priors show conjugate formulation and exact marginal likelihoods can be computed (Scott and Carvalho, 2008) but overall inflexible due to its restrictive forms. In the space of decomposable graph the marginal likelihood are available upto normalizing constants  (Giudici, 1996; Roverato, 2000). The marginal likelihoods are used to calculate the posterior probability of each graph, resulting an exact solution for smaller dimension but for a moderately large $P$(number of nodes) or outside such restrictive class the computation may be prohibitively expensive. For non decomposable graph the computation is non trivial and maybe prohibitive using  reversible jump MCMC (Giudici and Green, 1999; Brooks et al, 2003). A  novel Monte Carlo technique can be found in Atay-Kayis and Massam (2005). There have been approaches by shrinking the covariance matrix using matrix factorization. For example, factorization of covariance matrix  in terms of standard deviation and correlation (Barnard et al., 2000), decomposition of correlation matrix (Liechty et al., 2004) explore such technique. Writing the inverse covariance matrix as the product of inverse partial variance and  the matrix of partial correlations, Wong et al. (2003) used reversible-jump-based Markov chain Monte Carlo (MCMC) algorithms to identify the zeros among the off-diagonal elements.

An equivalent formulation of GGM is via neighborhood selection  through the conditional mean under normality assumption (Peng et al. 2009). The method is based on the conditional distribution of  each variable, conditioning on all other variables. In a GGM framework, this conditional distribution is a normal distribution with the conditional mean function linearly related to the other variables. Furthermore, the conditional independence relationship among variables can be inferred by the variable selection techniques of the regression coefficients of the conditional mean function (Meinshausen and B\"{u}hlmann (2006)). More specifically, if a specific  regression coefficient appeared to be zero, the corresponding variables are conditionally independent. Of course, the joint distribution approach and the conditional approach based on linear regressions are essentially equivalent.

\begin{figure}
\centering
\includegraphics[width=2in,height=2in]{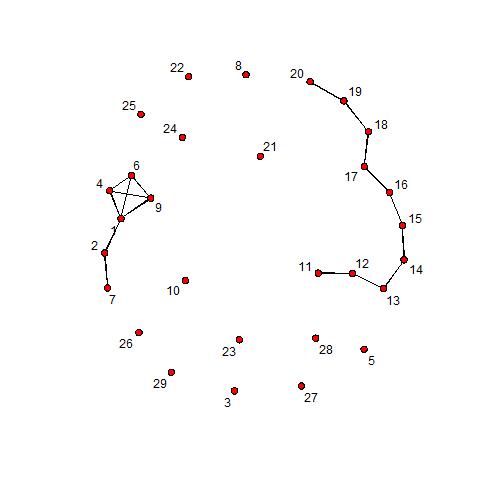}
\includegraphics[width=2in,height=2in]{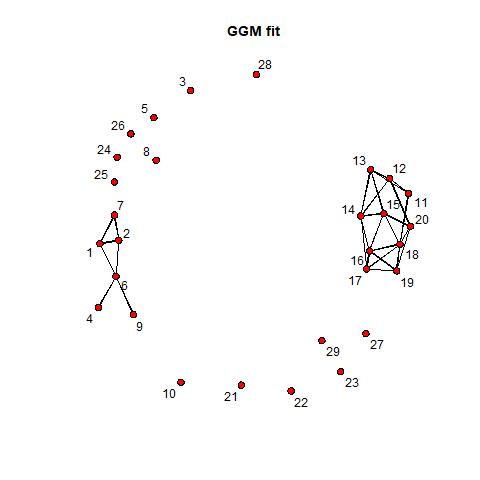}
\label{gaus_fig}
\caption{Left panel shows the true graph and  right panel shows GGM fit in a typical case.}
\end{figure}

Due to ease of computation and the presence of a nice interpretation, the vast majority of works on graphical model selection have been centered around the multivariate Gaussian distribution. In a multivariate Gaussian setup the conditional mean conveys necessary and sufficient information to infer the conditional independence structure. In contrast, for other distributions, this may not be true. For instance, for the multivariate t-distribution, the conditional independence can not be captured only using the conditional mean as it also depends on the conditional variance which is a nonlinear function of  other variables (Kotz and Nadarajah, 2004). For more complex distributions, the conditional independence structure may depend nonlinearly on higher order moments of the conditional distribution. Hence, the inference of a graph can be significantly affected by deviations from the normality and can lead to a wrong graph. The following example, which we discuss in details in  section 5 (Example 1 (a)), demonstrates the effect of deviation from normality in a simple case.  We assume the following  structure  for a graph with 30 variables/nodes $X_1,\dots,X_{30}$ with 400 observations from each variable. Here $X_{11},\dots,X_{20}$ is generated from a  heavy tailed distribution induced by a common scale parameter, and  $X_i, X_j$. $i,j\leq 10$ is connected in the network iff $|i-j|<2$, given the scale parameter,  and $X_{1}\dots,X_{9}$ has some nonlinearity and non-normality and they form a subgraph $G_1$ disjoint from $G_2$ formed by $X_{11},\dots,X_{20}$.  We have $X_{20},\dots,X_{29}$ independent of the rest and $X_{30}$ is the function of the latent scale parameter.  The fitted and true graphs for $X_1,\dots,X_{29}$ given the scale parameter, are given in Figure \ref{gaus_fig} where index $i$ denotes $i$ th vertex corresponding to  $X_i$, and  it is clear that with deviation from Gaussianity   we have a large number of falsely detected edges. 
 This poses serious restriction in a variety of applications which contain non-Gaussian data as well as data with outliers. Liu et al. (2012) used Gaussian Copula model to allow flexible marginal distributions. Alternatively,  non-Gaussian distributions have been directly used  for modeling the joint distribution to obtain the graph (Finegold and Drton, 2011; Yang et al., 2014).

   In this paper, we propose a novel  Bayesian quantile based graphical model. The main intention is to model the conditional quantile functions (rather than the mean) in a regression setup. This is well known that the conditional quantile regression coefficients can infer the conditional independence between variables.  Under linearity of the conditional quantile regression function,   conditional distribution of the $k$th variable is  independent of the $j$th variable if  the corresponding  regression coefficient of the quantile regression  is zero for all quantiles. Hence by performing a neighborhood selection of these quantile regression coefficients, we can explore the graphical structure. Thus, in our framework this  neighborhood selection boils down to a variable selection problem in the  quantile regression setup. A spike and slab prior  formulation has been used for that purpose (George and McCulloch, 1993).   The likelihood function depends on a grid of  quantiles and borrowing strength from several quantile regression parameters is allowed through a hierarchical Bayesian model. Using Bayesian approach through spike and slab type prior, we can characterize the uncertainty regarding selected graph through the posterior distribution. 

A natural development would be to investigate the asymptotic property of the proposed estimated graph.   We  study the asymptotic behaviors of the graph when  the dimension as well as the number of observations increases to infinity.  The posterior probability of a small Hellinger neighborhood around the true graph approaches to one,  under conditions similar to Jiang (2007). Subsequently,  we extend this proof of consistency under the assumption of model misspecification, even under heavy-tailed distribution with sub-exponential tail bound. 
%

 The posterior distribution is not in an explicit form, hence we  resort to simulation based MCMC method. However, carrying out  MCMC in this complex setup could be computationally intensive. Therefore along with MCMC, we also propose a variational algorithm for  the mean field approximation of the posterior density (Beal, 2003; Wand et al., 2011; Neville et al., 2014).  

The main contributions of our paper are: (1)  development of robust graphical models based on quantiles in a Bayesian hierarchical modeling framework,  (2) proving the consistency of those resultant graph estimates under  truly specified as well as misspecified models and, (3) proposing the MCMC based posterior simulation technique as well as a fast computationally efficient approximation of the posterior distribution. 

In  next section, we formulate the neighborhood selection problem for a particular node and write down the corresponding likelihood and the posterior density.  In section 3, we discuss the estimation consistency. Later in section 4, we discuss the posterior approximation in details and write down the network construction algorithm. In section 5, we discuss some of the examples and in section 6, we use the proposed method in establishing a protein network.

\section{Methodology}

An undirected graph $G$ can be represented by the pair $(V,E)$, where $V$ represents the set of {\it vertices} and $E=(i,j)$ represents the set of edges, for some $i,j \in V$.  Two nodes, $i$  and $j$, are called {\emph neighbors}  if $(i,j)\in E$. A graph is called {\it complete}, if all possible pair of nodes are {\emph neighbors}, $(i,j)\in E$ for every $i,j\in V$. $C \subset G$, is called {\emph complete} if it induces a complete subgraph.
A Gaussian graphical model (GGM) uses a graphical structure to define a set of pairwise conditional independence relationships on a $P$-dimensional constant mean, normally distributed random vector $\bx\sim N_{P}(\bmu,\bSigma_{G})$. Here $\bSigma_{G}$ denotes the dependence of the covariance matrix $\bSigma$ on the graph $G$ and this is the key difference of this class of models with the usual Gaussian models. Thus, if $G=(V,E)$ is an undirected graph and if $\bx=(x_{\nu})_{\nu\in V}$ is a random vector in $R^{|V|}$ that follows a multivariate normal distribution with mean vector $\bmu$ and covariance matrix $\bSigma_{G}$  then the unknown covariance matrix $\bSigma_{G}$ in GGM is restricted by its Markov properties; given $\bOmega_{G}={\bSigma_{G}}^{-1}$, elements $x_{i}$ and $x_{j}$ of the vector $\bx$ are conditionally independent, given their neighbors, iff $\omega_{ij}=0$ where $w_{ij}$ is the $ij$th element of $\Omega_{G}$. If $G=(V,E)$ is an undirected graph describing the joint distribution of $\bx$, $\omega_{ij}=0$ for all pairs $(i,j)\not\in E$. Thus, the elements of the adjacency matrix of the graph $ G$ have a very specific interpretation, in the sense that they model conditional independence among the components of the multivariate normal. Presence of an off-diagonal edge in the graph indicates non-zero correlation while its absence indicates zero correlation. This way, the covariance matrix $\Sigma$ (or the precision matrix $\Omega$) depends on the graph $G$ and this dependence is denoted as $\Sigma_{G}$ ($\Omega_{G})$. The equivalent results can be obtained by using the conditional regression setup where the conditional distribution of one variable $X_{k}$ given all other variables is $[X_{k}|X_{-k}]\sim N(\sum_{j\neq k}\beta_{kj}X_{j},\sigma_{k}^{2})$ where $\beta_{kj}=-\omega_{kj}/\omega_{kk}$, $\sigma_{k}^{2}=1/\omega_{kk}$ and $X_{-k}$ is the vector containing  all $X$s except the $k$th one. It is clear that the variable $X_{k}$ is conditionally independent of $X_{l}$ given all other variables iff the corresponding conditional regression coefficient $\beta_{kl}$ is 0. This result transforms the Gaussian graphical model problem to a variable selection (or neighborhood selection) problem in a conditional regression setup (Meinshausen and B\"{u}hlmann (2006)).

If  multivariate normality assumption on ${\bf x}$ does not hold, then the conditional mean does not characterize the dependence among the variables.    Under general distribution, it can be helpful to study  the full conditional distribution. The absence of an edge between $k$th and $j$th node implies that the conditional distribution of $X_k$ given the rest $X_k|X_{-k}$, does not depend on $j$ and vice versa. Any distribution is characterized by its quantiles. Therefore, we can look at the conditional quantile functions of $X_k$ and check if it depends on $X_j$. Hence, the main idea is to model the quantiles of $X_k$ and perform a variable selection over all quantiles. We use linear model for modeling the quantile functions and perform  variable selection in  the set up of quantile regression (Koenker and Bassett, 1978; Koenker, 2004) .

Thus, we generalize the concept of Gaussian graphical model in a quantile domain where we consider the conditional quantile regression  of  each of the node variable $X_k$ given all others say $X_{-k}$ for $k=1\cdots P$. In a conditional linear quantile regression model if $X_k (\tau)$ is the $\tau$ th quantile of  $k$th variable  $X_k$  then the conditional quantile of $X_{k}$ given $X_{-k}$, that is $X_{k,-k} (\tau)$,  can be expressed as 
\begin{equation}
X_{k,-k} (\tau)=\beta_{k,0}(\tau)+\sum_{j\neq k}\beta_{k,j}(\tau)X_j,\: j=1,\cdots P.
\label{qtile1}
\end{equation}
We summarize the above discussion in the following result.
\begin{proposition}
Under the assumption of linearity  of the conditional quantile function of $X_k$, as in  model \eqref{qtile1},      $X_k$ is conditionally independent of $X_j$ iff  $\beta_{k,j}(\tau)=0, \forall \tau$. 
\label{qtl_prp}
\end{proposition}
\noindent Therefore from Proposition \ref{qtl_prp}, we obtain a similar framework as in the Gaussian graphical model problem.  That way, we transform the quantile graphical modeling problem to a quantile regression problem.

   Furthermore, instead  of looking at a single quantile such as  median, considering a set of quantiles will be useful to address a more general dependence structure.   To induce sparsity, it will be helpful to look at the coefficients for a set of quantiles $\tau_s$ and assume that  the condition $\beta_{k,j}(\tau_s)=0$ for all $s$ implies the conditional independence among the corresponding variables.  Indeed, the sparse  graphical model  based on \eqref{qtile1} addresses  more general cases than just modeling the conditional mean. In practice  instead of the continuum, grid points $0<\tau_1<\dots<\tau_m<1$ are used for the selection process.

In many practical scenarios, conditional quantiles may not be linear over all quantiles and over all the variables. In that case, we consider  $\hat{L}(X_{k,-k}(\tau))=\hat{\beta}_{k,0}(\tau)+\sum_{j\neq k}\hat{\beta}_{k,j}(\tau)X_j$, the best linear approximation that  minimizes the expected quantile loss function $E[\rho_\tau(X_k-{L}(X_{k,-k}(\tau))]$ where $L(\cdot)$ varies over all linear functions and,  the quantile loss function is given by  $\rho_\tau(z)=z\tau, z\geq 0$; $\rho_\tau(z)=-(1-\tau)z , z<0$. We also assume that this minimizer is unique.  Next, we assume,

\begin{itemize}

 \item[C1.] If $X_{k,-k}(\tau)$ does not depend on $X_j$ for some $j$,  for any $\tau$, then the  coefficient of $X_j$ in $\hat{L}(X_{k,-k}(\tau))$ is zero over all quantiles,  that is  $\hat{\beta}_{k,j}(\tau)=0$ for all $\tau$; 
 
 \item[C2.]  If $X_{k,-k}(\tau)$ depends on $X_j$ for some $\tau$, then there exists $\epsilon,\delta, \delta'>0$  such that for $\tau$  on the interval  $[\epsilon, 1-\epsilon]$, we have  $|\hat{\beta}_{k,j}(\tau)|>\delta$ for $\tau$ in an open subset of $[\epsilon, 1-\epsilon]$ of radius $\delta'$ and $\hat{\beta}_{k,j}(\tau)$ is a continuous function of $\tau$ for $\tau \in (0,1)$.

\end{itemize}
\noindent Condition $C1$ enforces that conditional independence implies the same for best linear quantile function and condition $C2$ implies that $X_j$'s that are connected to a particular  $X_k$ are `detectable' through linear quantile regression. Condition $C2$ can be relaxed by using polynomial/spline basis to accommodate general functions, but here we restrict ourselves to linear functions and  linear quantile regression.

Suppose we have $n$ independent observations which can be presented as a $n\times P$ data matrix ${\bf X}^*=\{X_{ij},i=1,\cdots ,n,j=1,\cdots ,P\}$. We write  ${\bf X}^*=[X_{1},\cdots ,X_{P}]$  where $X_{i}$ is the $n\times 1$ dimensional $i$th column vector containing the data corresponding to the $i$th variable. Since we consider the conditional quantile regression for each of the  variable $X_{j}$ given all the other variables, for the sake of simplicity we describe the general methodology only for a specific variable $X_k$. For notational convenience, we assume $Y$ is the $k$th column of ${\bf X}^*$ containing the data related to $X_{k}$. Furthermore, ${\bf X}={\bf X_{-k}}^{*}$ is a ${n\times P}$ dimensional   matrix containing data corresponding to all other variables except the $k$th one. Hence, we redefine ${\bf X}$ having  $X_i$  in the $i+1$ th column if $ i<k$ and $X_i$   in the $i$th column for $i>k$. We also allow  the intercept term as a vector of ones in the first column. In the quantile regression for $X_{k}$, we treat $Y$ as the response and ${\bf X}$ as the covariates.  For the $\tau$ th quantile regression, we obtain the estimates of the regression coefficients by minimizing the loss function $l$ such as
$\text{min }_{\bf\beta}\sum_{i=1}^n\rho_\tau(y_i -\bfx_i'{\bf \beta})$ the regression coefficient vector  ${\bf \beta}=\{\beta_0,\beta_1,\dots,$ $\beta_{k-1},\beta_{k+1},\dots,\beta_P \}$, $y_{i}$ is  the $i$th element of $Y$ and $\bx_{i}$ is the $i$th row of ${\bf X}$. 

Mathematically minimizing this loss function $l$ is equivalent to maximizing $-l$ where ${\rm exp}(-l)$ is proportional to the likelihood function.
 This duality between a likelihood and loss, particularly viewing the loss
as the negative of the log-likelihood, is referred to in the Bayesian literature as a logarithmic scoring rule (see, for example, Bernardo (1979), page 688). Using loss function to construct likelihood may cause model misspecification.  Later we address the issue and show even under model misspecification, we have the posterior concentration around the best linear approximation of the conditional quantile functions.  
Accordingly, the corresponding likelihood based method can be formulated by developing the model as  $y_i=\bx_i'{\bf \beta}+u_i$ where $u_i$s are independent and identically distributed (iid) random variables with the scale parameter $t$ as $f(u|t)=t\tau(1-\tau)exp(-t\rho_\tau(u)).$

Using the likelihood  corresponding  to the quantile regression gives the consistent estimate of the coefficients of the conditional quantile regression (Sriram et al., 2013).  Misspecified likelihood  (see Chernozhukov and Hong, 2003; Yang et al., 2015) may impact the posterior inference  such as confidence interval for coefficients. But  here our main goal is  to model the conditional quantile function through linear approximation and perform a model selection for the quantile function through a likelihood equation.  Also, we do not enforce any ordering restriction between the quantile functions for different quantiles. If the linear representation holds for conditional quantile  then  the posterior estimates from the likelihood  based on the loss function should show the desired ordering, as we can estimate the coefficients of the quantile regressions consistently. 

The quantile based conditional distributions may not correspond to a joint distribution. However, here we  model the linear approximation of the conditional quantile functions over a grid of quantiles and construct posterior probability  of the selecting the neighbors of a particular node/variable by constructing the pseudo likelihood function based on quantile loss.  Later we show that even if we have misspecified model, we have posterior probability of selecting wrong edge/neighbor will go to zero under this  loss based pseudo likelihood.

Using the results from  Li et al.(2009)  and Kozumi and Kobayashi (2009), we can express
: $u_i=\xi_1v_i+t^{-\frac{1}{2}}\xi_2\sqrt{v_i}z_i$,
where $\xi_1=\frac{1-2\tau}{\tau(1-\tau)}$, $\xi_2=\sqrt{\frac{2}{\tau(1-\tau)}}$ , $v \sim Exp(t)$ and $z \sim N(0,1)$. Furthermore, the variables indexed by different $i$ s are independent. 

The final model can be represented by integrating previous results as
\begin{eqnarray}
y_ i&=&\bfx_i'{\bf \beta}+\xi_1v_i+\xi_2t^{-\frac{1}{2}}\sqrt{v_i}z_i \nonumber\\ 
v_i &\sim& Exp(t), \nonumber\\
z_i&\sim& N(0,1).
\label{qtlrp2}
\end{eqnarray}

For selecting the adjacent nodes (neighborhood selection) for  node $k$, a Bayesian variable selection technique has been  performed.  The stochastic search variable selection (SSVS) is adapted  using a spike and slab prior for the regression coefficients as : $ p(\beta_j|I_j)=I_jN(0,g^2v_0^2)+(1-I_j)N(0,v_0^2)$,  (George and McCulloch (1993))
for $j=1,\dots,P, j\neq k$ and $I_{j}$ is the indicator variable related to the inclusion of the $j$th variable.  Let $\gamma$ be the vector of indicator function $I_j$'s.  We denote the spike variance as $v_0^2$ and the slab variance as  $g^2v_0^2$, where $g$ is a large constant.   Alternatively, writing $\beta_{\gamma,j}=\beta_j I_{j}$ ( Kuo and Mallick, 1998) can be helpful, where we use the indicator function in the likelihood and model the quantile of $y$ by $  x'\beta_\gamma$.  Further, a Beta-Binomial prior is assigned for $I_j$.  The corresponding Bayesian hierarchical model  is described as
\begin{eqnarray}
\beta_j &\sim& N(0,t^{-1}\sigma_\beta^2),\nonumber\\ 
I_j &\sim& Ber(\pi),\nonumber\\ 
\pi  &\sim& Beta(a_1,b_1),\nonumber\\ 
t &\sim& Gamma(a_0,b_0).
\label{prior1}
\end{eqnarray}

The Beta Binomial prior opposed to a fixed binomial distribution with a fixed $\pi$  induces sparse selection (Scott and Berger, 2010).

For a sparse estimation problem we consider $m$ different  quantile grid  points in $(0,1)$ as $\tau_1\dots,  \tau_m$. Let $\underline{\bf \beta}_l=\{\beta_{0,l},.., \beta_{k-1,l},\beta_{k+1,l}\dots\beta_{P,l}\}$ be the coefficient vector corresponding to the $\tau_l$ quantile and $\underline{\bf \beta}_{\gamma,l}=\{\beta_{0,l},.., \beta_{k-1,l}I_{k-1,l},$ $\beta_{k+1,l}I_{k+1,l}, $  $\dots\beta_{P,l}I_{P,l}\}$.  Let $\underline{\bf \beta}$ be the vector of all the $\underline{\beta}_l$'s; $\underline{\mathbf \beta}_{(l-1)P+j}^{}=\beta_{j-1,l}^{}$ if $j<k$ and $\underline{\mathbf \beta}_{(l-1)P+j}^{}=\beta_{j,l}^{}$ for $j>k$. In this setup,  $I_{j,l}=0$ for all $l$ implies that $X_j$ is not in the model,  and $I_{j,l}=1$ for some $l$  implies that $X_j$ is included in the model. 
 Let $t_l$ be the scale parameter for $\tau_l$. For $\tau_l$, we write $v_i$, $\xi_1$ and $\xi_2$ from (\ref{qtlrp2}) as $v_{i,l}$, $\xi_{1,l}$ and $\xi_{2,l}$, respectively. Let ${\bf v}$ be the vector of $v_{i,l}$'s. Using $\tau_1,\tau_2,\dots,\tau_m$ the corresponding loss function for $\tau_l$   is $ \hspace{0.1in}l(\underline{{\bf \beta}}_l)=\rho_{\tau_l}(y_-\bfx'\underline{\bf \beta}_{\gamma,l})
$
and the corresponding likelihood function is
\begin{eqnarray}
f_{\tau_l}(y_i|t_l,\underline{\bf \beta}_l,\gamma) \propto t_l \exp(-t_l\rho_{\tau_l}(y_i -\bfx_i'\underline{\bf \beta}_{\gamma,l})).
\label{qtllikeli}
\end{eqnarray}
The hierarchical model can be written as follows:
\begin{eqnarray}
\underline{\beta}_{l}|\tau_l & \sim & MVN_P({\mathbf 0}_P,\Sigma_{\beta, P\times P}), l=1,\dots,m,\nonumber\\ 
I_{j,l}&\sim& Ber(\pi_l),\nonumber \\
\pi_l  &\sim& Beta(a_1,b_1),\nonumber\\ 
t_l    &\sim& Gamma(a_0,b_0),\nonumber\\
f_{\tau_l}({\bf Y}|t_l,\underline{\bf \beta}_l,\gamma)& \propto& t_l^n \exp(-t_l\sum_{i=1}^n\rho_{\tau_l}(y_i -\bfx_i'\underline{\bf \beta}_{\gamma,l})).
\label{hierar}
\end{eqnarray}
Here, ${\mathbf 0}_P$ is a vector of zeros of length $P$ and,  $MVN_P({\mathbf 0}_P,\Sigma_{\beta, P\times P})$ denotes $P$ dimensional  multivariate normal distribution with the mean vector ${\mathbf 0}_P$ and the covariance matrix  $\Sigma_{\beta, P\times P}$.  We use ${\Pi}(.)$ to denote prior distributions. 

 Using the  setting in \eqref{hierar} and (\ref{qtlrp2}), we can express the posterior distribution of the unknowns as
\begin{eqnarray}
\Pi_l(\underline {\bf \beta}_l,\{I_{j,l}\}_{j\neq k},\pi_l,{\bf v}_l,t_l|{\mathbf Y}) \propto t_l^{3n/2}  \{ \prod_{i=1}^n {v_{i,l}}^{-\frac{1}{2}}\exp(- t_l\frac{(y_ i-\bfx_i'\underline{\bf\beta}_{\gamma,l}-\xi_{1,l}v_{i,l})^2}{2v_{i,l}\xi_{2,l}^2})\times  \nonumber \\  exp(-t_lv_{i,l})\}  \Pi(\underline{\bf \beta}_l)\}\prod_{j \neq k}\Pi(I_{j,l})  \Pi(\pi_l) \Pi(t_l).
\label{posterior2}
\end{eqnarray}
%
%

 Each of the posteriors $\Pi_l(\cdot)$ gives probability to the parameters and hyper-parameters  corresponding to $\tau_l$ in particular, on ${\mathbf \Theta}_l=\{\underline {\bf \beta}_l,\{I_{j,l}\}_{j\neq k},\pi_l,{\bf v}_l,t_l\}$. Let, ${\mathbf \Theta}=\{{\mathbf \Theta}_l\}_l$.  The distribution on ${\mathbf \Theta}$ induced by $\Pi_l(\cdot)$'s given by   $\Pi(\mathbf{\Theta})=\prod\Pi_l(\mathbf{\Theta}_l)$. 

 The posterior distribution given in \ref{posterior2} is not available in an explicit form and we have to use simulation based approach like Markov Chain Monte Carlo (MCMC) to obtain realizations from it which is described in section 4. Even more, we have to repeat this procedure for each $k$ over all quantiles, which makes it more computationally demanding. Due to these reasons,  we also develop an approximate method based on the variational technique. 

\section{Graph estimation consistency}
In this section, we consider the consistency of the proposed graphical model. Two approaches can be adopted. One method is to look at the variable selection consistency for each of the nodes and the alternative way will be to consider the fitted density induced by the graphical model. We take the latter approach first and show the predictive consistency of the proposed network in scenarios encompassing the $P>n$ case.   The dimension is adaptively increased with increasing $n$, the number of observations for each variable. Let $P=p_n$ be the number of nodes. We show that with $n$ increasing to infinity under some appropriate conditions on the prior, the fitted density lies in the Hellinger ball of radius $\epsilon_n$, approaching to zero, around the true density with high probability, if the proposed model is correct.  Next, we consider the case of model miss-specification and neighborhood selection consistency.

\subsection{Consistency under true model}

 Convergence in  exponential rates in terms of Hellinger distance between the posterior graph and the true graph  can be achieved under conditions similar to Jiang (2007). Here we briefly define the convergence criterion, describing the conditions required and discuss their implications in terms of the  graph estimation.

 Let, $G^*$ be the true graph and $f_{k,G^*}$ be the density associated with the  $k$ the node of true graph under the proposed model and $\Pi(.|.)$ be the posterior density and $f_k$ be the density under the model given in Equation \ref{qtllikeli} . Convergence in terms of Hellinger distance such as,
\[ P_{G^*}[\Pi[d(f_k,f_{k,G^*})<\epsilon_n|{\bf X^*}]>1-\delta_n]\geq1-\lambda_n  \] where $\epsilon_n, \delta_n, \lambda_n$ going to zero as $n \rightarrow \infty$ for each $k$, can be achieved. Here, $d(f,f^*)=\sqrt{(\int_{\chi} (\sqrt{f(x)}-\sqrt{f^*(x)} )^2 d(\nu(x)})$ denotes the scaled standard Hellinger distance in  some measure space $\chi$ with measure $\nu$, where $f^*$ be the true data generating density, and $P_{G^*}[\cdot]$ or $P^*[\cdot]$ be the probability under true data generating density.

For the neighborhood of $Y=X_k$, writing the coefficients $\underline{\bf \beta}_{\gamma,l}^{(-k)}=\underline{\bf \beta}_{\gamma,l}$, $\underline{\bf \beta}_{l}^{(-k)}=\underline{\bf \beta}_{l}$, ${\beta}^{(-k)}_{j,l}={\beta}_{j,l}$, $v_{i,l,k}=v_{i,l}$, using \eqref{posterior2} we have
\begin{eqnarray}
\Pi_l(\underline {\bf \beta}^{(-k)}_l,\{I_{j,l}^{(-k)}\}_{j},\pi_l,{\bf v},t_l|.) \propto \hspace{0.5in} &&\\ \nonumber t_l^n  \big\{ t_l^{\frac{n}{2}}  \{  \prod_{i=1}^n {v_{i,l,k}}^{-\frac{1}{2}}\exp(- t_l\frac{(X_ {i,k}-{\bf{\bfx^*_{-k}}_i}'\underline{\bf\beta}_{\gamma,l}^{(-k)}-\xi_{1,l}v_{i,l,k})^2}{2v_{i,l}\xi_{2,l}^2})\times \\ \nonumber  exp(-t_lv_{i,l,k})\}  \Pi(\underline{\bf \beta}_l^{(-k)})\}\prod_{j \neq k}\Pi(I^{}_{j,l})  \Pi(\pi_l) \Pi(t_l)\big\}
\label{posterior3}
\end{eqnarray}
where ${\bfx^*_{-k}}_i$ is the $i$ th row of ${{\bf X}^*_{-k}}$.  Through this conditional modeling, we  show the posterior concentration of $f(x_k|x_i,i\neq k)f^*(x_i,i\neq k)$ around $f^*(x_1,\dots,x_P)$.

For the indicator function for the neighborhood selection of $k$ th node, we assume $I_{j,l}\sim Ber(\pi_n), j\neq k,$ with the restriction $\sum_{j=1}^{p_n}I_{j,l}\leq\bar{r}_n$. Let ${r}_n =p_n\pi_n$. The restriction on the maximum possible dimension can be relaxed by assuming a small probability on the set 
$\sum_{j=1}^{p_n}I_{j,l}\geq\bar{r}_n$. Also, the scale parameter $t_l=t$ is assumed to be fixed. The following results also hold for the Beta-Binomial prior on the indicator function   and we  address  it later. 

Let $\epsilon_n$  be a  positive sequence decreasing to zero and $1 \prec n\epsilon_n^2 $, where $a_n \prec b_n $ implies $\frac{b_n}{a_n} \rightarrow \infty$. 
  We have the following prior specifications, $\underline{ \bf \beta}^{(-k)} _l\sim MVN_{P}({\bf 0}_{P},S_{\beta_l}^{-1})$, where $S_{\beta_l}^{-1}$ is a diagonal matrix in our setting. 

Under the true data generating model given in Equation  \ref{qtllikeli} , let $\Delta^l_k(r_n)=inf_{|\gamma|={r}_n} \sum_{j \notin \gamma, j\neq k}|{\beta^*}^{(-k)}_{j,l}|$.  Here the superscript '$*$' denotes the true coefficient values. Let $ch_1(M)$ denote the largest eigenvalue of the some positive definite matrix $M$. Let $\underline{\beta}_{\gamma_l}^{(-k)} \sim N(0,V_{\gamma_l})$, i.e the distribution restricted to the variables included in the model. 
Let, $B(r_n)=max_l\{ch_1(V_{\gamma_l}), ch_1(V_{\gamma_l}^{-1})\}$. 

Suppose the following conditions hold.

A1. \hspace{.5in}$\bar{r}_nlog ( p_n) \prec n\epsilon_n^2$.

A2.  \hspace{.5in} $\bar{r}_nlog (1/ \epsilon_n^2) \prec n\epsilon_n^2$.

A3. \hspace{.5in} $1\leq r_n\leq \bar{r}_n\leq p_n$.

A4.  \hspace{.5in}$ \sum_{ j\neq k}|{\beta^*}^{(-k)}_{j,l}|<\infty$.

A5.  \hspace{.5in}$1\prec r_n \prec p_n <n^\alpha;\alpha>0$.

A6.  \hspace{.5in}$B(\bar{r}_n) \prec n\epsilon_n^2$.

A7.  \hspace{.5in}$p_n \Delta^l_k(r_n)\prec\epsilon_n^2$.

Conditions similar to A1-A7 can be found in Jiang (2007). Condition A1 is needed for establishing the entropy bound on a smaller restricted  model space, that is an upper bound on the  number of Hellinger balls needed to cover the restricted model space.  Conditions A2 and A6 ensure that we have sufficiently large prior probability on the Kullback-Leibler(KL) neighborhood of true model.  Assumption A7 is needed to ensure sparsity that is coefficients from all but few variables are close to zero and the total residual effect is small. Also, it has $p_n$ multiplied on the L.H.S as we may  not have the boundedness of the node values. Also, the eigenvalue condition is satisfied trivially. 

The main idea is to show negligible prior probability for models with dimension larger than $\bar{r}_n$ or where the coefficient vector lies outside a compact set. Then next step would be to cover the smaller model  space with $N(\epsilon_n)$ many  Hellinger balls of size $\epsilon_n$ with $log(N(\epsilon_n))\prec n\epsilon_n^2$. Tests can be constructed similar to Ghosal et al.(2000). Then by showing that  the prior probability of   KL neighborhood  around the true model has  lower bound of some appropriate order, the following results can be achieved. 


Let, 
 $h_k= \sqrt{(\int_{\chi} (\sqrt{f(x_k|x_i, i\neq k)}-\sqrt{f^*(x|x_i,i\neq k)} )^2 f^*(x_{i\neq k})d({\bf x})})$.  
  Let the generic term $D_n$ denotes the data matrix.
Then we have the following theorem.

\begin{thm}
Suppose $sup_j E|X_j|=M^*<\infty$ . Then from \eqref{posterior2}  under A1-A7,   for some  $c_1',c_2'>0$ and for $n^\delta \prec p_n\prec n^\alpha$; $\alpha>\delta>0$ and for $sup_{|\gamma_l|\leq\bar{r}_n}\{ch_1(V_{\gamma_l}), ch_1(V_{\gamma_l}^{-1})\}\leq B\bar{r}_n^v$; $v,B>0$, for large enough $\bar{r}_n$,  the following convergence results hold   in terms of the  Hellinger distance if the true data is generated by the likelihood given by equation \eqref{qtllikeli} for some $\tau_l$, as the number of observations goes to infinity.

$a)$
\begin{eqnarray*}
P^*[\Pi_l(h_k\leq \epsilon_n|D_n)>1-e^{-c_1'n\epsilon_n^2}] \rightarrow 1.
\end{eqnarray*}


%
\begin{proof}
Given in the Appendix section.
\end{proof}
\label{consthm1}
\end{thm}

\begin{rmk}
In particular,  for $\bar{r}_n \prec n^b$ with $b=min\{\xi,\delta,\xi/v \}$ and $\epsilon_n=n^{-(1-\xi)/2}$ with $\xi \in (0,1)$, we have $n\epsilon_n^2=n^\xi$ and the convergence rate of the order $e^{-n^\xi}$.

\end{rmk}
\begin{rmk}
If each of the node has finitely many neighbors, then some assumptions  on tail conditions such as A4, A7 become redundant as only finitely many ${\beta^*}^{(-k)}_{j,l}$'s are non zero for each $k$. For $p_n=O(n^\alpha), 0<\alpha<1$, we can have $\bar{r}_n=p_n$ and $\epsilon_n=n^{-(1-\xi)/2}$, with $\xi\in (\alpha,1)$. Thus, we do not need to add any restriction on the model size. 
\end{rmk}

\begin{rmk}
The results in  Theorem \ref{consthm1}  hold for Beta-Binomial prior on the indicator function as well and given in the Appendix section.\end{rmk}

\subsection{Consistency under model misspecification}

\subsubsection{Density estimation}
Model \eqref{qtllikeli} has been developed from a loss function and   may not be the true data generating model. Therefore, we extend our consistency results under the condition of model misspecification. Let, $f^0_{k,-k}$ be the true density of $Y=X_k$ given ${\bf X_{-k}^*}$ and ${\mathscr{ F}}_k$ be the set of densities $f_{l,k,-k}$'s given by \eqref{qtllikeli}. Let $f^0_{-k}$ be the true data generating density for $X_{-k}$, the variables other that than $X_k$.  Let $f^*_{l,k,-k} \in {\mathscr{F}}_k$ be the density in  \eqref{qtllikeli} such that $f^*_{l,k,-k}f^0_{-k}$  has the smallest Kullback-Leibler (KL) distance with $f^0_{k,-k}f^0_{-k}$. We show that the  posterior given in \eqref{posterior2} concentrates around  $f^*_{l,k,-k}$ for $\tau_l$  .  We fix the scale parameter $t_l$.

Let $l_{\tau,\underline{\bf \beta}_l^{(-k)}}=E ( \rho_{\tau_l}(x_{k}-x_{-k}^*\underline{\bf \beta}_l^{(-k)}))$ and  $\hat{\underline{\bf \beta}}_l^{(-k)}= \text{arg min}_{\underline{\bf \beta}_l}l_{\tau,\underline{\bf \beta}_l^{(-k)}}$ and suppose the minimizers are unique.  Let, $\hat{\underline{\bf \beta}}^{(-k)}$  be the combined vector, analogous to ${\underline{\bf \beta}}^{(-k)}$. Then under some conditions, the posterior converges to $f_{l,k,-k}(\hat{\underline{\bf \beta}_l}^{(-k)}  )$, the density corresponding to the best linear quantile approximation for $\tau_l$. Let $\text{inf }KL( f^0_{k,-k}f^0_{-k},f_{l,k,-k}({\underline{\bf \beta}_l}^{(-k)} )f^0_{-k})=\delta_{k,l}^*$, which is achieved at the parameter value $\hat{\underline{\bf \beta}}^{(-k)}_l$.

Posterior concentration under model misspecification needs more involved calculations and can be shown under carefully constructed test functions, as given in Kleijn and Van der Vaart (2006  ) .  However,  such approach may depend on the convexity or boundedness of the model space. We take a route similar to Sriram et al. (2013) based on the quantile loss function  and show the  convergence directly.  To prove the consistency, we make a few assumptions. Without loss of generality, we assume that the variables are centered around zero. 

Let $d_k$ be the degree (the number of neighbors)  of the $k$ th node.   Under the following conditions we prove the convergence theorem.
\begin{itemize}
\item[]B1. max$_kd_k<M_0-1 $ for some universal constant $M_0$.

\item[]B2. $E(e^{\lambda |X_k-E(X_k)|})\leq e^{.5\lambda^2\nu^2} \text{ for } |\lambda|<b^{-1}, \forall k,\nu>0$ (sub-exponential tail condition).

\item[]B3. There exists $\epsilon>0$,  such that for $|X_k|<\epsilon, \forall k$, any $m\leq M_0$ dimensional  joint density of  any $m$ number of covariates $X_k$'s is uniformly bounded away from zero. We also assume that $X_k$'s have uniformly bounded second moments. 
\item[]B4. $log p_n \prec n$.
\item[]B5. $sup_k\|\hat{\underline{\bf \beta}}^{(-k)}\|_\infty<\infty$.
\end{itemize}
%
%

\begin{thm}
From \eqref{qtllikeli} and \eqref{posterior2}, under conditions B1--B5, for any $\delta>0$, $\Pi(KL( f^0_{k,-k}f^0_{-k},f_{l,k,-k}f^0_{-k})$  $>\delta+\delta_{k,l}^*|.) $ goes to zero almost surely for all  $ k$, as the number of observations goes to infinity.
\label{thm_mis}
\end{thm}
Next, we derive the posterior convergence rate under the model misspecification. For a sequence $\epsilon_n$ converging to zero, we assume 
\begin{itemize}

\item[]B6. $\epsilon_n \sim n^{-\xi}, \xi<.25$, 
\item[]B7.  $logp_n \prec n\epsilon_n^4$.
\end{itemize}

\begin{thm}
From \eqref{qtllikeli} and \eqref{posterior2}, under conditions B1--B7,  as the number of observations goes to infinity, $\Pi(KL( f^0_{k,-k}f^0_{-k},f_{l,k,-k}f^0_{-k})$ $>\delta_n+\delta_{k,l}^*,$ $ \text{ for some }k) $ goes to zero almost surely,  where $\delta_n =4 \epsilon_n^2$. 
\label{mis_rate}
\end{thm}
Proofs of Theorems \ref{thm_mis} and \ref{mis_rate} are given in the  Appendix section. We first show the results for bounded $X_k$'s and later extend our results for heavy-tailed sub-exponential distributions, at the end of the proof of Theorem \ref{mis_rate}.

\subsubsection{Neighborhood selection consistency}

Next, we state the following Theorems about the neighborhood selection. For $X_k$ or $k$ th node, let ${\it N}_k^*=\Large\{i\neq k;i\in\{1,\dots,P=p_n\}: X_i\leftrightarrow X_k\Large\}$,  where $X_j\leftrightarrow X_k$ implies that there is an edge between $j$ th and $k$ th node. Let, ${\it N}_{l,k}^*=\{i\neq k;i\in\{1,\dots,P=p_n\}: \hat{\beta}_{k,i}(\tau_l)\neq 0\}$, be the neighborhood corresponding to best linear conditional quantile for $\tau_l$, where $\hat{\beta}_{k,i}(\tau_l)$'s are given in conditions $C1,C2$ in Section 2 and $\hat{\beta}_{k,i}(\tau_l)$ is the coefficient corresponding to $X_i,i\neq k$, in $\hat{\underline{\beta}}^{(-k)}_l$ from Section 3.2.1.

\begin{lem}
Under $C_1$ and $C_2$, for $0=\tau_0<\tau_1<\tau_2\dots<\tau_m<1$ and $\tau_{i}-\tau_{i-1}<\delta_i$, there exists $\delta_0,m_0>0$ such that  for  $\delta_i<\delta_0$ for all $i$, $m>m_0$ and ${\it N}_k^*=\cup_l {\it N}_{l,k}^*$.
\label{lem1}
\end{lem}
Let $M_{l,k}^*$ be the model corresponding to the neighborhood ${\it N}_{l,k}^*$ and $M^*_k$ corresponds to ${\it N}_k^*$.

We assume the following.
\begin{itemize}

\item[]B8. $I_{j,l}\sim Ber(\pi_n)$ and  $-log\pi_n=O(n^{0.5+\epsilon'}); 0<\epsilon'<0.5$.
\item[]B9. $logp_n=O(log n)$.
 \end{itemize}
 
 The above condition $B8$ puts a strong penalty on the model size which penalizes the neighborhood size of a node, and selection probability  under posterior distribution of any bigger model, containing the true model for a node,  goes to zero with high probability.

 Next,  we assume the following for the conditional densities and the quantiles. This conditions are similar to the conditions in Angrist et al. (2006) in the context of estimating the conditional quantile regression coefficient for miss-specified linearity.  For a model $M^1_k$ at node $k$, let $Z$ denote the $|M^1_k|+1$ dimensional random variable consisting of 1 in the first place and $X_j$'s,  $j\neq k$ that are in the model $M^1_k$  in the remaining places and $|M^1_k|$ is the size of model $M^1_k$.  Let $\hat{\underline{\beta}}(\tau)_{M^1_k}$ be the corresponding coefficients for the best linear conditional quantile.

 \begin{itemize}

\item[C3.] The true conditional density $f^0(x_k|x_{-k})$ is bounded and uniformly continuous  in $x_k$ uniformly over support of $X_{-k}$.
\item[C4] $J(\tau)= E[f^0(Z'\hat{\underline{\beta}}(\tau)_{M^1_k}|Z)ZZ']$ is positive definite and finite for all $\tau$, for $Z$ defined above for any $M^1_k$,  and $E[\|Z\|^{2+\epsilon_2}]$ is uniformly bounded for some $\epsilon_2>0$, over all possible model of size $|M^1_K|$, for any finite dimensional  model $M^1_k$.
 \end{itemize}

Let $\hat{\underline{\beta}}(\tau_l)$ the coefficient vector that minimizes the linear conditional quantile regression loss $E[\rho_{\tau_l}(Y-\underline{\beta}'X)]$ where $Y=X_k$, and $\hat{\underline{\beta}}(\tau_l)_n$ be the MLE for  the likelihood based on this loss function.  In  Angrist et al. (2006),  convergence    of the process $\sqrt{n}(\hat{\underline{\beta}}(\tau)-\underline{\hat{\beta}}(\tau)_n)$ was shown, for $\tau$ in an open subset of $(0,1)$. Using those results we show the following neighborhood selection related result.
 
 Let, $\Pi^n_{\tau_l,k}(M_1,M_2)=\Pi^n_{l,k}(M_1,M_2)$ denote  the  ratio of posterior probabilities of model $M_1$ and $M_2$,  at node $k$ for $\tau_l$ based on $n$ observations. Let $M^*_{l,k}$ be model based on  the  neighborhood $N^*_{l,k}$ for $\tau_l$, and for a model  $M_1$,  corresponding to some node, let $|M_1|$ be its size or number of covariates/neighbors in the model. We assume $t_l$ is fixed and equal to one, without loss of generality.

\begin{thm}
For  quantiles $\{\tau_1,\dots,\tau_m\}$,  $\delta_i=\tau_i-\tau_{i-1}$, for equation \eqref{posterior2}, under $B1, B2, B5,$ $ B8, B9, C1-C4$  we have $sup_l \{\Pi_{l,k}^n(M^{1}_k,M^*_{l,k}): M^{1}_k \neq M^*_{l,k} \} \rightarrow 0$, in probability, for any alternative model $M_k^1$, as $n$ goes to infinity.
\label{bf1}
\end{thm}

%
%
%

\begin{rmk}

Let $M^1_k$ be any model corresponding to a neighborhood at node $k$,  $N^1_k$, which does not contain $N_k^*$. Let $\hat{\underline{\beta}}(\tau)_{M^1_k}$ be the corresponding minimizer of the expected linear conditional  quantile loss for that model. Suppose,  we assume  $\hat{\underline{\beta}}(\tau)_{M^1_k}$ to be continuous on $\tau$ and $\text{inf}_{\tau\in(\epsilon,1-\epsilon)}l_{\tau,\underline{\hat{ \beta}}_{M^1_k}}-l_{\tau,\underline{\hat{ \beta}}_{M^*_k}}>0$ for any $\epsilon>0$, where  $l_{\tau,\underline{\hat{ \beta}}_{M^1_k}}=E [\rho_{\tau}(x_{k}-Z'\underline{\hat{ \beta}}(\tau)_{M^1_k})]$ and  $Z$ is the $|M_k^1|+1$ dimensional random variable with one in the first coordinate and variables corresponding to ${M^1_k}$ in others.  Then under the set up of Theorem \ref{bf1}, we have $sup_{\tau\in(\epsilon,1-\epsilon)} \{\Pi_{\tau,k}^n(M_k,M^*_{k}): M_k \neq M^*_{k} \} \rightarrow 0$, in probability.  

Therefore heuristically, for large $n$, choosing quantiles on  $[\tau_\epsilon,1-\tau_\epsilon]$, $\tau_\epsilon>0$, even if we choose quantile densely, the false discovery rate should not keep on increasing with the number of quantile grids, and should stabilize. This conclusion is later verified in our simulation.
\label{unif_bf}
\end{rmk}
%


\section{Posterior analysis}
We first describe the  MCMC steps for  posterior simulation.  Next, we derive  the variational approximation algorithm steps for our case.  For simplicity, we illustrate the posterior sampling for  $Y=X_{k,n\times 1}$ and ${\bf X}={\bf X_{-k}}^{*}$; i.e the neighborhood selection for the $k$ th node.  For notational convenience, we will not use the suffix $k$ in this section and formulate the method for a regression setup.  Let us introduce some notations which will be used in both formulations.

Let ${\bf X}_{\gamma,l}$ is ${n\times P}$ dimensional  covariate matrix containing  $X_i {\bf I}_{i,l}$  in $i+1$ th column for $ i<k$ and $X_i {\bf I}_{i,l}$   in $i$th column for $i>k$ and  the vector of ones in the first column.

To write the steps for variational approximation and MCMC, we  define the following quantities.
\begin{itemize}
\item  Let $Y_1={\bf 1}_{m \times 1}  \otimes Y$ be an $n\times m$ length vector formed by replicating $Y, m$ times. 

\item Let ${\bf X}_{1,\gamma}$ be the matrix by arranging the ${\bf X}_{\gamma}$'s diagonally.  Let,  ${\bf X}_{1,\gamma}^E$ denotes the matrix ${\bf X}_{1,\gamma}$, with indicators replaced by their expectations. Similarly, we have ${\bf X}_{\gamma}^E$.

\item Let $Y_{(l-1)n+i}^\delta={Y_1}_{{(l-1)n+i}}-\xi_{1,l}(E(\frac{1}{v_{i,l}}))^{-1}$ for $l=1,\dots,m$. Similarly, let  $Y_{(l-1)n+i}^{\delta'}={Y_1}_{{(l-1)n+i}}-\xi_{1,l}(\frac{1}{v_{i,l}})^{-1}$, and $Y_{}^{\delta,l}, Y_{}^{\delta',l}$ be the analogous $n$ length vectors for $\tau_l$.

\item  Let $\Sigma_l$  be the $n \times n$ diagonal matrix, where $i$ th diagonal entry is $E(t_l)E(\frac{1}{v_{i,l}\xi_{2,l}^2})$ for $l=1,\dots,m$ . Similarly,  $\Sigma_l^1$  be the $n \times n$ diagonal matrix, where $i$ th diagonal entry is $(t_l)(\frac{1}{v_{i,l}\xi_{2,l}^2})$ for $l=1,\dots,m$ .

\item  Let $S_x={\bf X_1}'\Sigma{\bf X_1}$ and $S_{x,\gamma}={\bf X}_{1,\gamma}'\Sigma{\bf X}_{1,\gamma}$. Similarly, $S_{x,\gamma}^E$ is the expectation of $S_{x,\gamma}$. Let  $S_{x,\gamma,l}$ and  $S_{x,\gamma,l}^E$ be the matrices corresponding to $\tau_l$.
%

\item Let $\underline{ \bf \beta}$ be the $mP$ length vector such that $\underline{\mathbf \beta}_{(l-1)P+j}=\beta_{j-1,l}$ if $j<k$ and $\underline{\mathbf \beta}_{(l-1)P+j}=\beta_{j,l}$ otherwise, for $l=1,\dots,m$.  Also, note that we denote the prior for $\underline{ \bf \beta}$ as $\underline{ \bf \beta}\sim N(0,S_\beta^{-1})$ as in Section 3.
\end{itemize} 

\subsection{MCMC steps}

Here, we describe the implementation of the MCMC algorithm to draw realizations from the posterior distribution. More specifically, we use Gibbs sampling by simulating from the complete conditional distributions which are described below (for the $k$'th node).

\vspace{0.4in}
\noindent \underline{(a) For  the coefficient vector $\underline{\bf \beta}_l$}:

 Given rest of the parameters the conditional distribution is:
\[q^{new}(\underline{\bf \beta}_l|.) := MVN((S_{x,\gamma,l}+S_{\beta,l})^{-1}({\bf X}_{\gamma,l})'\Sigma_l^1 Y^{\delta',l},(S_{x,\gamma,l}+S_{\beta,l})^{-1}).\]
  $\Sigma_l^1$  be the $n \times n$ diagonal matrix, where $i$ th diagonal entry is $t_l(\frac{1}{v_{i,l}\xi_{2,l}^2})$ for $l=1,\dots,m$ and $S_{x,\gamma,l}={\bf X}'_{\gamma,l}\Sigma_l^1{\bf X}_{\gamma,l}$.  

\noindent \underline{(b) For  $\pi_l$ }:

 \[q^{new}( \pi_l|. ) := Beta(a_1+\sum_{j=1,j\neq k}^{P}  I_{j,l},P-1-\sum_{j=1,j \neq k}^{P} I_{j,l}+b_1).\]

\noindent \underline{(c) For  $v_{i,l}$'s:} 
 \[q^{new}(v_{i,j}|.)\propto v_{i,l}f_{InG}(v_{i,l},\lambda_{i,l},\mu_{i,l})\]
where 
$f_{InG}(v_{i,j},\lambda_{i,l},\mu_{i,l})$ is a Inverse Gaussian density with parameters $\lambda_{i,l}=t_l(\frac{(y_ i-\bfx_i'\underline{\bf\beta}_{\gamma,l})^2}{\xi_{2,l}^2})$ and $\mu_{i,l}=\sqrt{\frac{\lambda_{i,l}}{2t_l+t_l\frac{\xi_{1,l}^2}{\xi_{2,l}^2}}}.$
This step involves a further Metropolis-Hastings sampling with a proposal density for $v_{i,l}$'s as $f_{InG}(v_{i,j},\lambda_{i,l},\mu_{i,l})$.

\noindent \underline{(d) For $t_l$}:

\[q^{new}(t_l|.) := Gamma(a_2,b_2)\] where $a_2=a_0+\frac{n}{2}+n$ and $b_2=b_0+\frac{1}{2}\sum_{i}(\frac{(y_ i-\bfx_i'\underline{\bf\beta}_{\gamma,l}-\xi_{1,l}v_{i,l})^2}{v_{i,l}\xi_{2,l}^2})+\sum_{i} v_{i,l}$.
 
\:\:

\noindent \underline{(e) For the indicator functions}:

\[ \log(\frac{ P(I_{j,l}=1|.)}{ P(I_{j,l}=0|.)}) =(\log\frac{\pi_l}{1-\pi_l}) -\frac{1}{2}\{\sum_{{i},I_{j,l}=1}t_l(\frac{(y_ i-\bfx_i'\underline{\bf\beta}_{\gamma,l}-\xi_{1,l}v_{i,l})^2}{v_{i,l}\xi_{2,l}^2}) - \] \[\sum_{{i},I_{j,l}=0}t_l(\frac{(y_ i-\bfx_i'\underline{\bf\beta}_{\gamma,l}-\xi_{1,l}v_{i,l})^2}{v_{i,l}\xi_{2,l}^2})\}.\]

We simulate from this conditional distributions iteratively to obtain the realizations from the joint posterior distribution.

\subsection{Variational approximation}
As explained in section 2 , we approximate the posterior distribution to facilitate a faster algorithm. We use the variational Bayes methodology for this approximation. First, we briefly review the variational approximation method for posterior estimation. For observed data $Y$ with parameter ${\bf \Theta}$ and prior  $\Pi({\bf \Theta})$ on it, if we have a joint distribution  ${p(Y,\Theta)}$ and a posterior ${\Pi({\bf\Theta}|Y)}$ respectively then 
\begin{eqnarray*}
\log p(Y)&=&\int \log\frac{p(Y,\Theta)}{\Pi({\bf\Theta}|Y)} q({\bf \Theta}) d({\bf \Theta})\\
&=&\int \log\frac{p(Y,\Theta)}{q({\bf \Theta})} q({\bf \Theta}) d({\bf \Theta})+KL(q({\bf \Theta}),\Pi({\bf\Theta}|Y)),
\end{eqnarray*}
for any density $ q({\bf \Theta})$. Here $KL(p,q)=E_p(log\frac{p}{q})$, the Kullback-Leibler distance between $p$ and $q$. 
Thus, 
\begin{eqnarray}
\label{vrkl}
\log p(Y)&=&KL(q({\bf \Theta}),\Pi({\bf\Theta}|Y))+\mathcal{L}(q({\bf \Theta}),p(Y,{\bf\Theta})) \nonumber \\
 -\mathcal{L}(q({\bf \Theta}),p(Y,{\bf\Theta})) &= &  KL(q({\bf \Theta}),\Pi({\bf\Theta}|Y))-\log p(Y) . 
\end{eqnarray}
With given $Y$, we minimize  $ -\mathcal{L}(q({\bf \Theta}),p(Y,{\bf\Theta}))=\int log \frac{q({\bf\Theta})}{p(Y,{\bf \Theta})}q({\bf \Theta})d{\bf \Theta}$.  
Minimization of the L.H.S of \eqref{vrkl} analytically may not be possible in general and therefore, to simplify the problem, it is assumed that the parts of ${\bf \Theta}$ are conditionally independent given $Y$. That is
\[ q({\bf \Theta})=\prod_{i=1}^sq_i(\Theta_i)\]
and $\cup_{i=1}^s \Theta_i={\bf \Theta}$ is a partition of the set of parameters ${\bf \Theta}$. Minimizing under the separability assumption, an approximation of the posterior distribution is computed.  Under this assumption of minimizing L.H.S of \eqref{vrkl} with respect to $q_i(\Theta_i)$, and  keeping the other $q_j( . ), j\neq i$ fixed, we develop the  following mean field approximation equation:

\begin{eqnarray} q_i(\Theta_i) \propto exp( E_{-i}\log(p(Y,{ \bf \Theta})),
\label{vrupdate}
\end{eqnarray}
where $E_{-i}$ denotes the expectation with respect to $q_{-i}({\bf \Theta})=\prod_{j=1,j\neq i}^s $ $ q_j(\Theta_j)$. We keep on updating $q_i(.)$'s sequentially until convergence.

For $\tau_l$, we have   $  {\bf \Theta}={\bf \Theta}_l =\{  \underline{\bf \beta}_l, {\bf I}=\{I_{j,l}\},\pi_l,t_l, {\bf v}=\{v_{i,l}\}  \}$ with $i=1,\dots,n; j=1,\dots P, j \neq k. $   To proceed, we assume that the posterior distributions of $\underline{\bf \beta}_l, {\bf I}_l=\{I_{j,l}\},\pi_l, {\bf v}_l=\{v_{i,l}\} $ and $t_l$'s are independent given $Y$. Hence,
\[q({\bf \Theta}) =q(t_l)q(\underline{\bf \beta}_l)\prod_{j\neq k}q({ I_{j,l}})\prod_{i}q({ v_{i,l}})q(\pi_l) . \]

\begin{table}
    \caption{\label{table1}Variational Update Algorithm}
\resizebox{\columnwidth}{!}{
\begin{tabular}{|p{14cm}|}
\hline 
\vskip 2pt
1.  Set the initial values  $q^0(\underline{\bf \beta}_l), q^0({\bf v}_l)$,  $q^0({\bf I}_l)$, $q^0({t_l})$ and $q^0(\pi_l)$. We denote the current density by $q^{old}( )$.\\
{\bf For} iteration in 1:N  :

2.  Find $q^{new}(\underline {\bf \beta}_l)$ by
\begin{equation}
q^{new}(\underline {\bf \beta}_l)= \arg \mathop{\min}_{q^{*}(\underline {\bf \beta}_l)}{-\mathcal{L}}\left(q^{*}(\underline {\bf \beta}_l) q^{old}({\bf I}_l)q^{old}({\bf v}_l)q^{old}({\pi_l})q^{old}({t_l}),p({Y},{\bf \Theta}_l)\right ). \nonumber
\end{equation}
Initialize $q^{old}(\underline {\bf \beta}_l)=q^{new}(\underline {\bf \beta}_l)$.

3.  Find $q^{new}({\bf I}_l)$ by
\begin{equation}
q^{new}( {\bf I}_l)= \arg \mathop{\min}_{q^{*}( {\bf I}_l)}{-\mathcal{L}}\left(q^{old}(\underline {\bf \beta}_l)q^{*}({\bf I}_l)q^{old}({\bf v}_l)q^{old}({\pi_l})q^{old}({t_l}),p({Y},{\bf \Theta}_l)\right ). \nonumber
\end{equation}
Initialize $q^{old}({\bf I}_l)=q^{new}( {\bf I}_l)$.

4. Find $q^{new}( {\bf v}_l)$ by
\begin{equation}
q^{new}( {\bf v}_l)= \arg \mathop{\min}_{q^{*}( {\bf v}_l)}{-\mathcal{L}}\left(q^{old}(\underline {\bf \beta}_l)q^{old}({\bf I}_l)q^{*}({\bf v}_l)q^{old}({\pi_l})q^{old}({t_l}),p({Y},{\bf \Theta}_l)\right ). \nonumber
\end{equation}
We initialize $q^{old}({\bf v}_l)=q^{new}( {\bf v}_l)$.

5. Find $q^{new}( {\pi}_l)$ by
\begin{equation}
q^{new}( {\pi}_l)= \arg \mathop{\min}_{q^{*}( {\pi}_l)}{-\mathcal{L}}\left(q^{old}(\underline {\bf \beta}_l)q^{old}({\bf I}_l)q^{old}({\bf v}_l)q^{*}({\pi_l})q^{old}({t_l}),p({Y},{\bf \Theta}_l)\right ).\nonumber
\end{equation}
Initialize $q^{old}({\pi}_l)=q^{new}( {\pi}_l)$.

6. Find $q^{new}( {t}_l)$ by
\begin{equation}
q^{new}( {t}_l)= \arg \mathop{\min}_{q^{*}({t}_l)}{-\mathcal{L}}\left(q^{old}(\underline {\bf \beta}_l)q^{old}( {\bf I}_l)q^{old}({\bf v}_l)q^{old}({\pi_l})q^{*}({t_l}),p({Y},{\bf \Theta}_l)\right ). \nonumber
\end{equation}
Initialize $q^{old}( {t}_l)=q^{new}( {t}_l)$.

We continue until the stop criterion is met.\\
{\bf end for}\\
9. Return the approximation $q^{old}(\underline{\bf \beta}_l)q^{old}({\bf I}_l)q^{old}({\bf v}_l)q^{old}(\pi_l)q^{old}(t_l)$.\\
\hline 
\end{tabular}
}

\end{table}

Under \eqref{prior1}, \eqref{posterior2},  we have,
\begin{eqnarray}
p(Y,{\bf \Theta}) \propto& t_l^n  \{  t_l^{\frac{n}{2}}\{ \prod_{i=1}^n{v_{i,l}}^{-\frac{1}{2}} \exp(- t_l\frac{(y_ i-\bfx_i'\underline{\bf\beta}_{\gamma,l}-\xi_{1,l}v_{i,l})^2}{2v_{i,l}\xi_{2,l}^2}) exp(-t_lv_{i,l}) \}\}  \vspace{-.1in}\nonumber\\
& \times exp(-\frac{1}{2}(\underline{\bf \beta}_l' (S_{\beta_l})\underline{\bf \beta}_l))\pi^{\sum_{j\neq k} I_{j,l} +a_1}(1-\pi_l)^{P-1-\sum_{j\neq k} I_{j,l}+b_1}t_l^{a_0-1}exp(-b_0t_l).\nonumber \\
\label{likeli2}
\end{eqnarray}
Using the expression given in \eqref{vrupdate}, we have the variational algorithm given in Table 1. 

The densities under this variational approximation algorithm  converge very fast and that makes the algorithm many time faster than the standard MCMC algorithms. From \eqref{vrupdate} we have an explicit form of $q^{new} (.)$ and for our case  the updations inside the algorithm are given next.

\subsubsection{ Sequential updates}

If $q^{old}(\pi_l), q^{old}(t_l),q^{old}(v_{i,l}), q^{old}(I_j)$  are the proposed posteriors of $\pi_l,\{t_l\},v_{i,l}$ and $I_{j,l}$'s at  the current step of iteration, we update
\[ q^{new}(\underline{\bf \beta}_l) \propto exp( E^{old}_{\pi_l,t_l,v_{i,l},I_{j,l};i,j} log(p(Y,{\bf \Theta}) )) \] where $ E^{old}_{\pi,t_l,v_{i,l},I_j;i,j}$ denotes the expectation with respect to the joint density given by  $q^{old}(\pi_l) q^{old}(t_l)q^{old}\prod_{i}q^{old}(v_{i,l})$ $\prod_{j}q^{old}(I_{j,l})$.

We have the following closed form expression for updating the densities sequentially. At each step, the expectations are computed with respect to the current density function. 

Thus, for  the coefficient vector writing the update across $l$ quantiles:

\[q^{new}(\underline{\bf \beta}_l) := MVN((S_{x,\gamma,l}^E+S_{\beta_l})^{-1}({\bf X}_{\gamma,l}^E)'\Sigma_l Y^{\delta,l},(S_{x,\gamma,l}^E+S_{\beta_l})^{-1}).\]

For  $\pi_l$ we have, 
\[q^{new}( \pi _l) := Beta(a_1+\sum_{j=1,j\neq k}^{P} E( I_{j,l}),P-1-\sum_{j=1,j \neq k}^{P}E( I_{j,l})+b_1).\]

For  $v_{i,l}$'s 
 \[q^{new}(v_{i,j})\propto v_{i,l}f_{InG}(v_{i,l},\lambda_{i,l},\mu_{i,l})\]
where 
$f_{InG}(v_{i,j},\lambda_{i,l},\mu_{i,l})$ is a Inverse Gaussian density with parameters $\lambda_{i,l}=E(t_l)E(\frac{(y_ i-\bfx_i'\underline{\bf\beta}_{\gamma,l})^2}{\xi_{2,l}^2})$ and $\mu_{i,l}=\sqrt{\frac{\lambda_{i,l}}{2E(t_l)+E(t_l)\frac{\xi_{1,l}^2}{\xi_{2,l}^2}}}.$


%
For the indicator function we have

\[ \log(\frac{ P(I_{j,l}=1)}{ P(I_{j,l}=0)}) = E(\log\frac{\pi_l}{1-\pi_l}) -\frac{1}{2}\{\sum_{{i},I_{j,l}=1}E(t_l)E(\frac{(y_ i-\bfx_i'\underline{\bf\beta}_{\gamma,l}-\xi_{1,l}v_{i,l})^2}{v_{i,l}\xi_{2,l}^2}) - \] \[\sum_{{i},I_{j,l}=0}E(t_l)E(\frac{(y_ i-\bfx_i'\underline{\bf\beta}_{\gamma,l}-\xi_{1,l}v_{i,l})^2}{v_{i,l}\xi_{2,l}^2})\}.\]

For the tuning parameter  $t_l$,

\[q^{new}(t_l) := Gamma(a_2,b_2)\] where $a_2=a_0+\frac{n}{2}+n$ and $b_2=b_0+\frac{1}{2}\sum_{i}E(\frac{(y_ i-\bfx_i'\underline{\bf\beta}_{\gamma,l}-\xi_{1,l}v_{i,l})^2}{v_{i,l}\xi_{2}^2})+\sum_{i} E(v_{i,l})$.

All the moment computations in our algorithm involve standard class of densities. Hence, moments can be explicitly calculated and used in the variational approximation algorithm. Later in the examples we standardize the data and use $t_l=1$.

\subsection{ Algorithm for graph  construction   }

Let $A$ be the $P\times P$  adjacency matrix of the target graphical model. Fixing $ \tau_1,\dots,\tau_m$,  for $k=1,\dots,P$, we compute the posterior neighborhood for each node as follows:
\begin{itemize}
\item Construct $Y=X_k$ and ${\bf X_{-k}}^{*}$ as in section 2.
\item Compute the posterior of $\Pi({\bf \Theta}_l|Y)$,  by using MCMC or the Variational algorithm, where  $ {\bf \Theta}_l =\{  \underline{\bf \beta}_l, {\bf I}_l=\{I_{j,l}\},\pi_l,t_l, {\bf v}_l=\{v_{i,l}\}  \}$ with $i=1,\dots,n; j=1,\dots P, j \neq k$  for all $l$.
\item As mentioned earlier in Section 2,   $I_{j,l}=0$ for all $l$, implies that $\beta_{j,l}$ is not in the model,  and $I_{j,l}=1$ for some $l$  implies that they are included in the model for some $l$.  If $P(I_{j,l}=1|Y)>0.5$ for some $l$, then $A(j,k)=1$, and $A(j,k)=0$ otherwise.
\end{itemize}
Two nodes $i$ and $j$ are connected if at least one of the two is in the neighborhood of the other according to the adjacency matrix $A$.

\section{Some  illustrative examples}
In this section, we consider three simulation settings to illustrate the application of the proposed methodology. We compare our methodology with the neighborhood selection method for Gaussian graphical model (GGM), using the R package 'huge'(Zhao et al., 2012) where the model is selected by `huge.select' function. We considered graphical Lasso (GLASSO) for graph estimation.
We use $a_1=1, b_1=1$ in the Beta-Binomial prior. Using the setting of \eqref{prior1} and \eqref{likeli2}, we use an independent  mean zero Normal  prior on the components of $\underline{\beta}$. 

\subsection{Example 1}
\subsubsection*{Example 1(a)}

To illustrate our method, we consider the following example. We consider $P=30$ variables $X_1,\dots,X_{30}$. 
We construct $X_{1},\dots,X_{10}$ in the following sequential manner:
\begin{eqnarray*}
X_{1},\dots,X_{10}& \sim& Gamma(1,.1)-10,\\
X_{2}&=&.4X_{1}+\epsilon_{1},\\
X_{6}&=&1.1X_{1}+4X_{4}+1.3X_{9}+\epsilon_{2},\\
X_{7}&=&\Phi^{-1}(\frac{exp(X_{2})}{1+exp(X_{2})})+\epsilon_{3},
\end{eqnarray*}
where $\epsilon_{1}\sim N(0,1)$, $\epsilon_{2} \sim .5N(2,1)+.5N(-2,1)$, $\epsilon_{3} \sim N(0,1)$ and they are independent and independent of the $X_i$'s for each step. The quantity $\Phi$ denotes the cdf of standard normal distribution.

%
%
%
%
%
%

Next, we construct $X_{11},\dots X_{20}$ from hierarchical multivariate normal random variables $Y_1,\dots,Y_{10}$. That is for $i$ th observation,  $y'_{1,i},\dots,y'_{10,i}\sim MVN({\bf 0},\Sigma)$, $x_{j,i}=y_{j,i}r_{i}$, $j=1,\dots,10$,  with $\bf{0}$ is a vector of zeros and  $\Sigma_{kl}=.7^{|k-l|}$ and $\frac{1}{r_{i}}\sim Gamma(3,3)$, $i=1,\dots,n$,  $r_i$'s are  independent,  and we have $X_{10+j}=Y_j, j\geq 1$.
We generate independent normal random variables with mean zero and variance 1 for $X_{21}$ till $X_{29}$ and $X_{30}$ be the vector of $log (r_i)$'s. Hence, we have total number of nodes/variables $P=30$, and given the scale parameter $X_{30}$, the graph has two disjoint parts namely: $\it{G}_1=\{X_1,\dots,X_9\}$ and $\it{G}_2=\{X_{11},\dots,X_{20}\} $. 
In addition, non-linear relationships are present between the variables.

 Generating $ n=400$ independent observations over 100 replications, we construct the network by our algorithm and compare it with the GGM  based neighborhood selection  method as mentioned earlier.  For GGM we use {\it `huge.select'} from the {\bf R} package {\it `huge'} which uses GLASSO and the implementations of the formulation from   Meinshausen and  Buhlmann (2006) (MB). The stability based selection criterion (argument{\it `stars'} in the {\bf R} function) performs  relatively better in this example and is therefore compared with our method.

\begin{figure}
\centering
\includegraphics[width=1.5in,height=2in]{true_ex1.jpg}
\includegraphics[width=1.5in,height=2in]{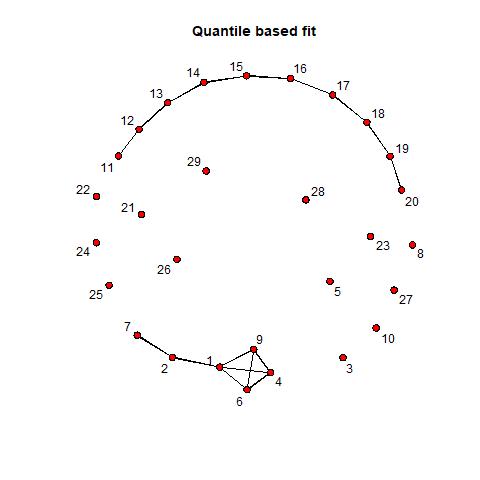}
\includegraphics[width=1.5in,height=2in]{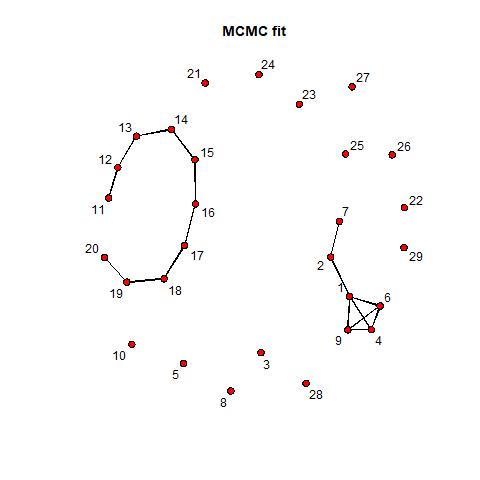}\\
\includegraphics[width=1.5in,height=2in]{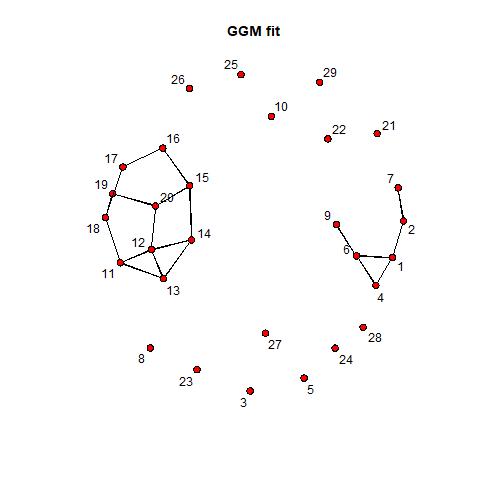}
\includegraphics[width=1.5in,height=2in]{ggm2_ex1.jpg}

\caption[Example]{Example 1(a). Network for $X_1,\dots,X_{29}$, conditional on the scale parameter. Top left panel shows the true subgraph. Top middle and right panel  show the network constructed by the quantile based variational Bayes (QVB)  and MCMC with   $\{\tau\}=\{0.3,0.5,0.7\}$, respectively.

 Bottom  right and left panel show  constructions by GGM based method using GLASSO and MB in {\it huge.select}, respectively. Index $i$ denotes $i$ th vertex corresponding to $X_i$. }
\label{qtlvsglasso}
\end{figure}

 For  quantile based variational Bayes (QVB),  the data is standardized, and we use  $t=1$, independent $N(0,v)$, $v=1$ prior on the coefficients. The QVB graph is robust to prior variance over a range $1\leq v\leq 100$.  A typical fitted subgraph for $X_1,\dots,X_{29}$ conditional on the scale parameter $X_{30}$  is presented in Figure \ref{qtlvsglasso} for QVB and MCMC based fits with same parameter specifications. The   QVB method has successfully recovered the connected part inducing sparsity whereas GGM has  estimated  wrong connections.  Moreover, the quantile based method performs better to separate the independent parts. Using MCMC algorithm, we obtain the similar graphs but the QVB is several hundred times faster than the  MCMC.  In Table \ref{tab2}, an account of false positivity ( detecting an edge, where there is none) has been provided  along with the average number of undetected edges for the QVB.
  Here, $FDR$ denotes the number of falsely detected edges on average per graph, $e_1$, $e_2$ and $e_3$ denote the average number of undetected edges in $\it{G}_1$, $\it{G}_2$ and  the average number of falsely detected connectors between them.  It can be seen that the misspecifications are significantly higher in GGM. The GGM detects a lot of extra edges along with the existing edges.   Also, $\it{G}_1$ and $\it{G}_2$ are generally well separated by the quantile based method.   Overall, the quantile based variational Bayes provides a sparser and a more accurate solution.   A typical MCMC fit is similar to    QVB fit (Figure \ref{qtlvsglasso})   but MCMC fits generally have slightly sparser graph with QVB detecting weaker connections more frequently. 

\begin{figure}
\centering
\includegraphics[width=2.5in,height=2.5in]{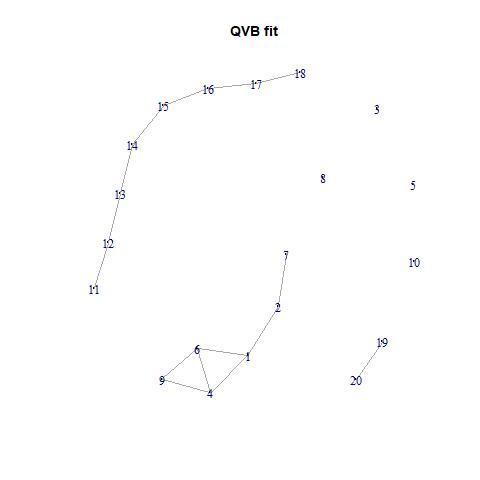}
\includegraphics[width=3.6in,height=2.5in]{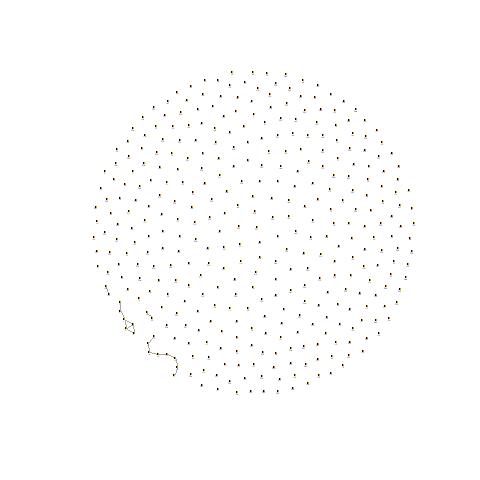}
\caption[Example]{Right panel shows a sparse fitted graph by variational Bayes for Example 1 $P>n$ case ($P=370, n=350$), with $\{\tau\}=\{.5\}$  and  left panel shows the connected variables. 
}
%
 
\label{diag}
\end{figure}
\begin{table}
\caption{\label{tab2}A comparison between GGM and quantile based variational Bayes method(QVB) for example 1. }
\centering
\begin{tabular}{| c | c    c   c   c| }
\hline
Method  & FDR & $e_1$ &$e_2$ & $e_3$  \\
\hline
$QVB(\{\tau\}=\{.5\})$ &0.28  &0.32 &0&0   \\

$QVB(\{\tau\}=\{.3,.5,.7\})$&0.58  &0.17 &0 &0  \\
\hline
GGM(MB)&5.21& 1.52 &0.07 &.08 \\
GGM(GLASSO)&14.78 &3.09&0.03  &.02\\

\hline
 \end{tabular}
\end{table}

 \subsubsection*{Example 1(b):
$P>n$ case. }

In the next example, we consider a sparse $P>n$ scenario. We construct $X_1,\dots,X_{10}$ similar to Example 1 (a). Next, we construct $X_{11},\dots X_{20}$ from a similar hierarchical multivariate normal random variables $Y_1,\dots,Y_{10}$. That is $y'_{1,i},\dots,y'_{10,i}\sim MVN({\bf 0},\Sigma)$, $x_{j,i}=y_{j,i}r_{j,i}$, $j=1,\dots,10$,  with $\bf{0}$ is a vector of zeros and  $\Sigma_{kl}=.7^{|k-l|}$ and $\frac{1}{r_{j,i}}\sim Gamma(3,3)$, $i=1,\dots,n$, $r_{j,i}$'s are independent,   and we have  $X_{10+j}=Y_j, j\geq 3$ and
\begin{eqnarray*}
X_{11}&=&3Y_3+2Y_5+\epsilon_4,\\
X_{12}&=& 3Y_6+2Y_7+\epsilon_5,\\
X_{10+j},&=&Y_j, j\geq 3. 
\end{eqnarray*}
\noindent Like the previous setup  of Example 1(a) with $n=350$ and  with adding further noise variables $X_{21},\dots,X_{370}$ which are generated from a standard normal distribution. Thus, we have  $n=350$ and $P=370$. The data is standardized and we use the same setting as of Example 1(a).
The proposed method performs well to detect the underlying latent structure, as well as provides a sparse solution (see Figure \ref{diag}).
%
%

%

\subsection{Example 2: Performance under Gaussianity}
Here, we compare  quantile based method with the GGM based methods, where the true data is Gaussian.  First we construct simple structured graph such as  hub-graph and  band graph (with banded structure in inverse covariance and adjacency matrix), and then generate multivariate normal data matrices with those underlying structures. We  use quantile based fit and compare with GGM based fit for such Gaussian data. For the next example, we consider sparse graphs.   The parameter specification for quantile based variational Bayes (QVB) is similar to that of last example.

\subsubsection{Hub-graph and Band graph} 
\begin{figure}
\centering
\includegraphics[width=1.6in,height=1.6in]{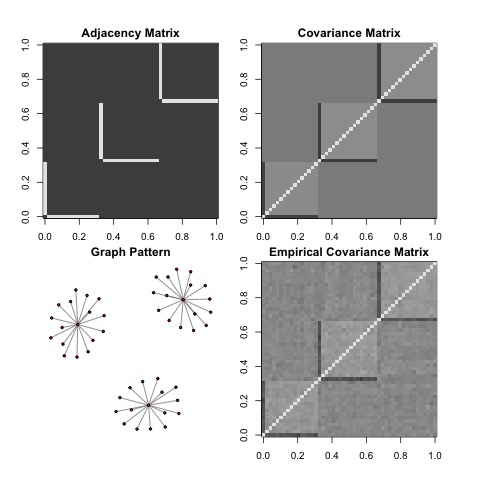}
\includegraphics[width=1.6in,height=1.6in]{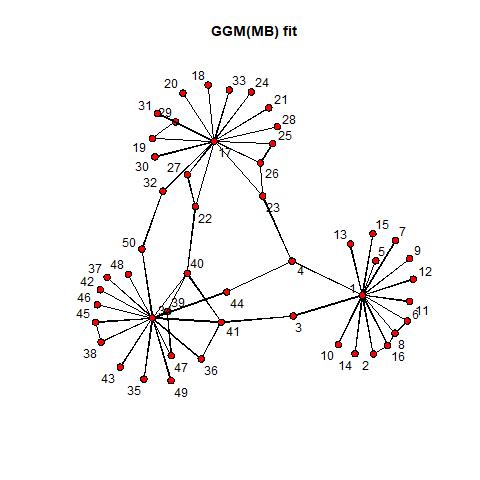}
\includegraphics[width=1.6in,height=1.6in]{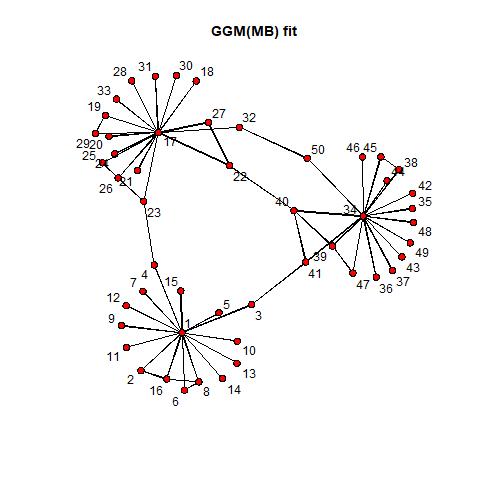}\\
\includegraphics[width=1.6in,height=1.6in]{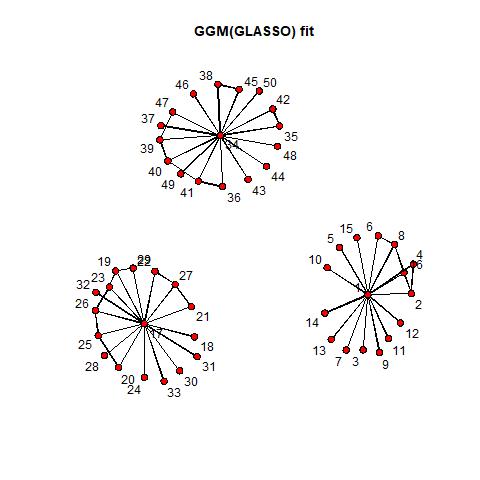}
\includegraphics[width=1.6in,height=1.6in]{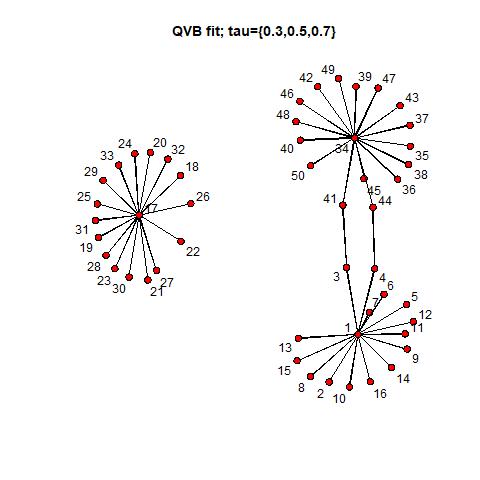}
\includegraphics[width=1.6in,height=1.6in]{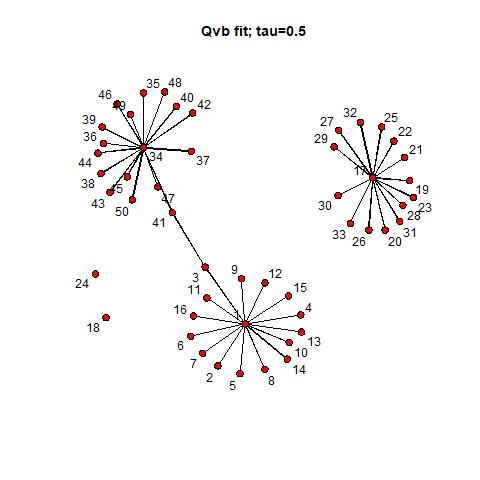}\\
\includegraphics[width=1.6in,height=1.6in]{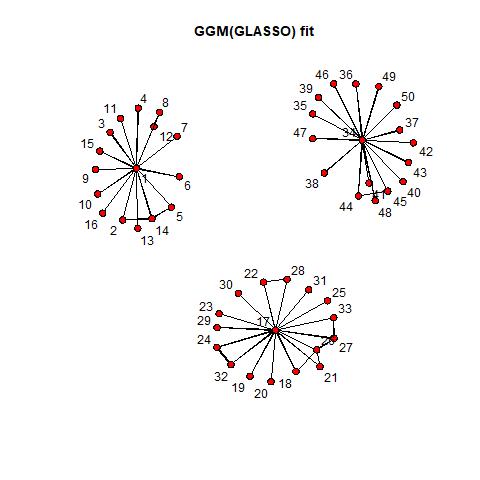}
\includegraphics[width=1.6in,height=1.6in]{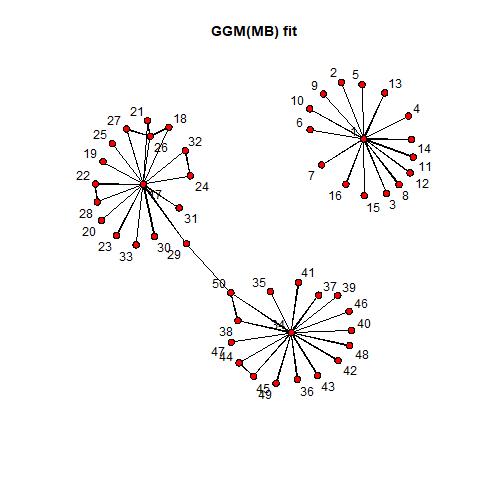}
\includegraphics[width=1.6in,height=1.6in]{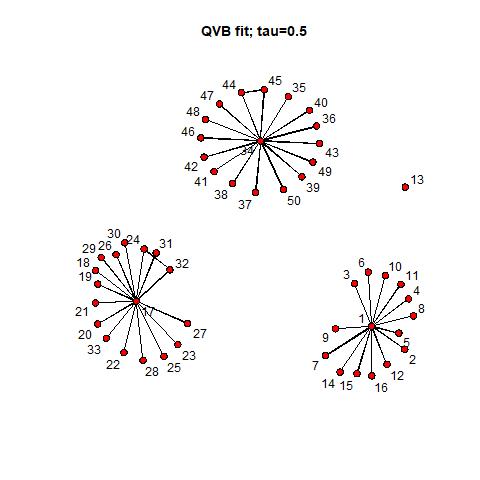}
\caption[Example]{Example 2, part 1. Upper row, left panel shows the true and generated quantities for hub-graph through {\it huge.generate}. Upper row,  middle and right panel show the GGM(MB) fit for stability and information based selection criterion, respectively. Middle row  shows the fit for GLASSO and the fitted  graphs for QVB. 

Bottom row shows the fit for another replication  with same set up. Here the fitted networks for $\tau=0.5$ and $\tau=\{0.3,0.5,0.7\}$ are same for QVB, and GGM selection criterions are stability based. }
\label{hub1}
\end{figure}

Using $n=300, P=50$ and $3$ hubs, we generate hub graph using {\it  huge.generate} function. A typical generated graph, with adjacency and inverse covariance given in Figure \ref{hub1} along with the GGM fit.  Here the nodes correspond to $1,\dots,50$ are  $X_1,\dots,X_{50}$ and the hub centers are located at $X_1,X_{17}$ and $X_{34}$. From the fitted graphs for $\tau=\{0.5\},\{0.3,0.5,0.7\}$ for QVB in Figure \ref{hub1}, it is evident that quantile based method's performance is  similar to GGM based methods, with QVB resulting slightly sparser graphs. 

\begin{figure}
\centering
\includegraphics[width=2in,height=1.6in]{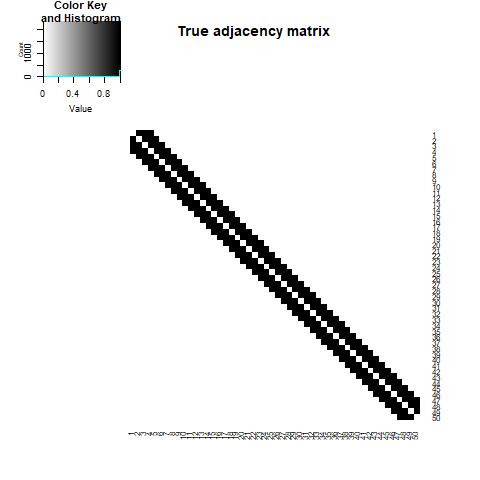}
\includegraphics[width=2in,height=1.6in]{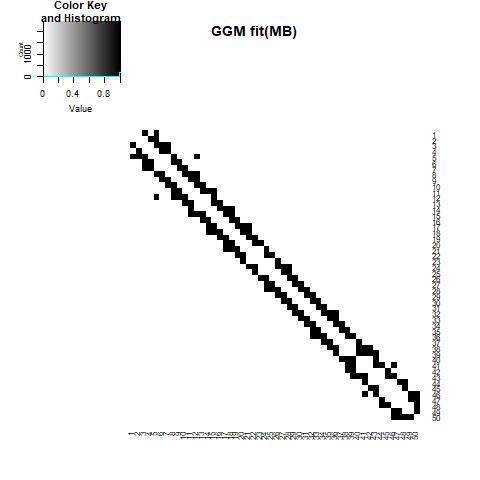}\\
\includegraphics[width=1.7in,height=1.6in]{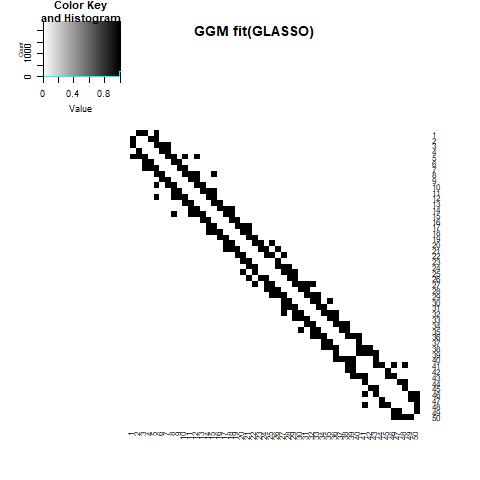}
\includegraphics[width=1.7in,height=1.6in]{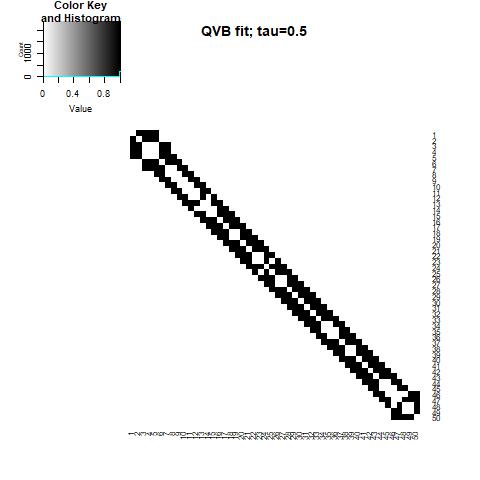}
\includegraphics[width=1.7in,height=1.6in]{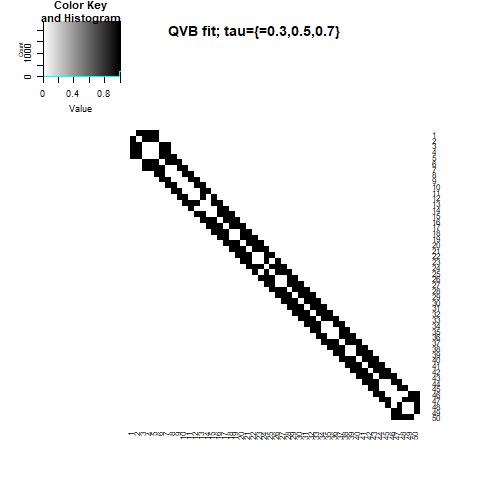}
\caption[Example]{Example 2, part 1. Upper  panel shows the true adjacency  matrix and  the GGM (MB) fitted adjacency for band graph, in left and right panel. Lower left,middle and right panel show the adjacency matrix for the GGM fit (GLASSO), QVB fit for $\tau=\{0.5\}$, and QVB for $\tau=\{0.3,0.5,0.7\}$, respectively. Index $i$ denotes $i$ th vertex corresponding to $X_i$. In the adjacency matrix $(i,j)$ th place is given by black iff $X_i\leftrightarrow X_j$ or the corresponding value is 1, and otherwise  given by white for zero or no edge. }
\label{band1}
\end{figure}

Next, we generate graph with underline covariance matrix having a band structure with $n=300$ and $P=50$, with nodes/covariates $X_1,\dots,X_{50}$,  where for $|i-j|\leq3$ there is an edge between $X_i$ and $X_j$. The  fitted and true adjacency matrices are  given in Figure \ref{band1}, where  the QVB's performance   compares favorably   to that of  GGM's. 

\subsubsection{Sparse Gaussian graph}
\begin{figure}
\centering
\includegraphics[width=2in,height=2in]{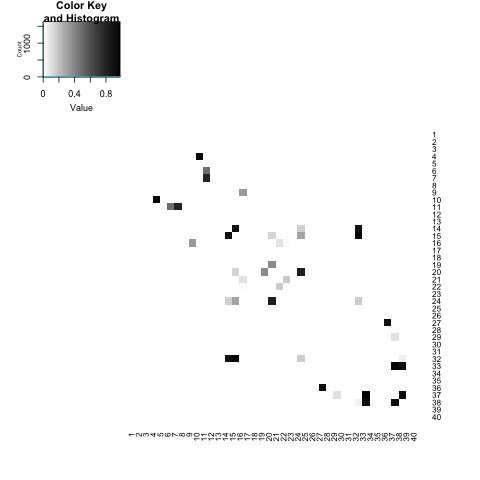}
\includegraphics[width=2in,height=2in]{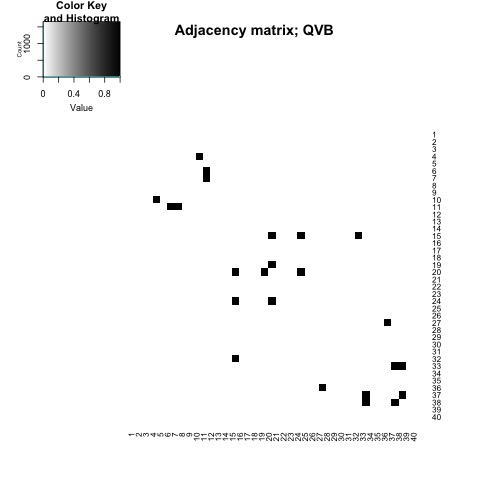}
\includegraphics[width=2in,height=2in]{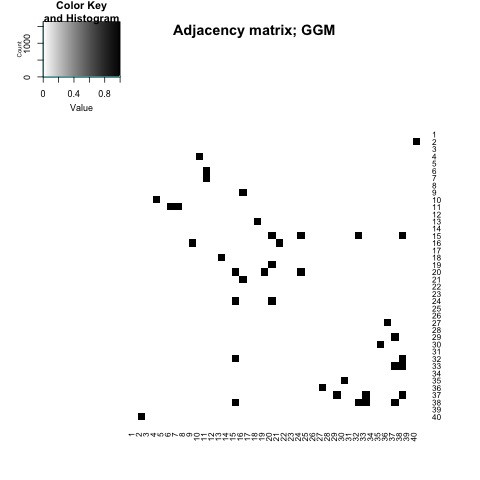}
\caption[Example]{Example 2, sparse graph case.  Here $n=100, P=40$, and sparsity level $=0.05$. Left panel shows the absolute value of partial correlation between variables, when data is generated from Gaussian graphical model. Middle and right panel shows the fitted adjacency matrices for QVB and GGM, respectively. We have $\tau=\{0.5\},\{0.3,0.5,0.7\}$ with both resulting same adjacency matrix for QVB.   Index $i$ denotes $i$ th vertex corresponding to $X_i$. In the adjacency matrix $(i,j)$ th place is given by black iff $X_i\leftrightarrow X_j$ or the corresponding value is 1, and otherwise  given by white for zero or no edge. }
\label{gaus1}
\end{figure}

\begin{figure}
\centering
\includegraphics[width=2in,height=2in]{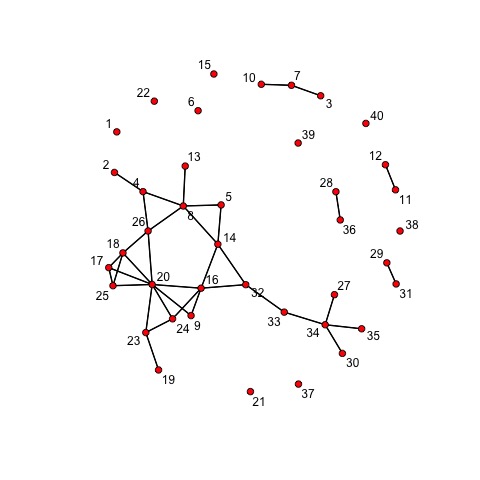}
\includegraphics[width=2in,height=2in]{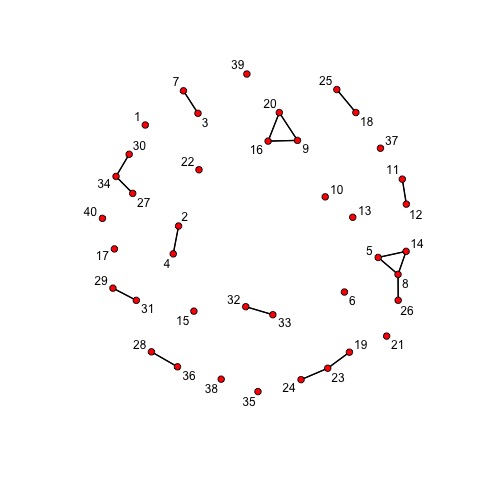}
\includegraphics[width=2in,height=2in]{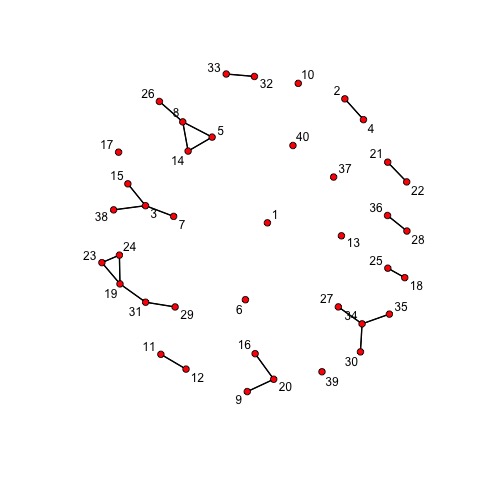}\\
\includegraphics[width=3in,height=2in]{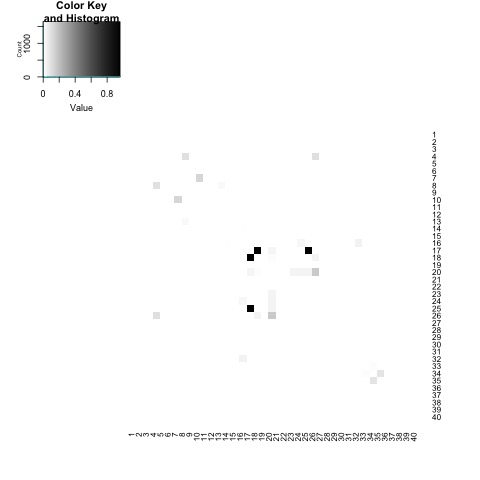}
\includegraphics[width=3in,height=2in]{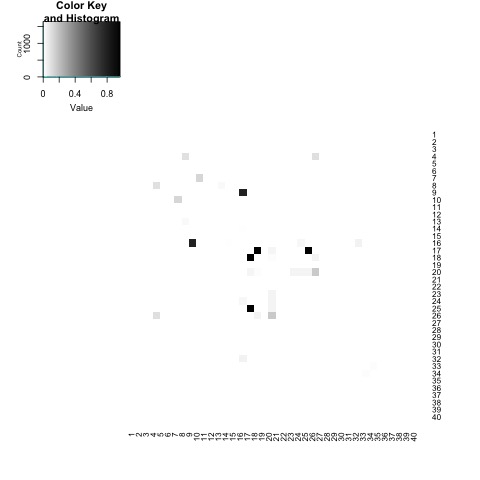}
\caption[Example]{Example 2, sparse Gaussian graph. Top left panel shows the true graph for the Gaussian graphical model.Top  middle and right panel shows the fitted graphs for QVB and GGM, respectively. Here $n=100, P=40$, and sparsity level $=0.1$. Bottom panel  shows the absolute value of the  partial correlations corresponding to the undetected connections,  for  QVB and GGM, in left and right panels, respectively. Mostly weaker connections have not been detected both in GGM and QVB.}
\label{gaus2}
\end{figure}

We generate graphs for different sparsity levels using the $R$ function {\it simulategraph}  and compare the quantile based fit with the GGM fit. Here, $P=40,n=100$ and sparsity levels are $.05,.1$ and thus, we have nodes corresponding to $X_1,\dots,X_P$.  Figure  \ref{gaus1} shows  the matrix of absolute values of the true partial correlation for the underlying true covariance matrix, for  the sparsity level $0.05$ and the corresponding  adjacency matrices of the  fitted network by QVB method for $\tau=\{0.5\},\{0.3,0.5,0.7\}$ and the GGM based fitted graph. The partial correlation is zero if and only if there is no edge between corresponding indices. The strength of the edge is proportional to the magnitude of this  partial correlation. It can be seen that QVB results in a sparse graph similar to GGM.
Generally QVB generates a sparse graph where very weak connections may not  be detected, similar to GGM based method. Figure  \ref{gaus2} shows a case with sparsity level $0.1$ where the partial correlation values for the  most of the undetected edges are close to zero and we have a sparse graph where the  relatively stronger connections are detected in both cases.  We use $MB$ specification in GGM with default  information criterion({\it ric}) based selection, which performs relatively better in this example. 

\subsection{Example 3: Effect of quantiles and computational gain}
\subsubsection{Example a.  Detecting the effect on extreme values}
 {\bf Example 3.a.i.} Next, we consider the case where the conditional distribution of one variable depends on the other in extreme values. 
For $X_1,\dots,X_{15}$ independent normal with mean zero and variance one,   $X'_{1}=2|X_{4}|+1.5|X_{7}|+.5N(0,1)$ and $X'_{2}=1.5|X_{5}|+2|X_{8}|+.5N(0,1)$. Let, $Z_{1}=X'_{1}{\bf I}_{X'_{1}>6}+N(-2,1){\bf I}_{X'_{1}<6}$ and $Z_{2}=X'_{2}{\bf I}_{X'_{2}>5.5}+N(-2,1){\bf I}_{X'_{2}<5.5}$, and $Z_p=|X_{p}|,p=3\dots,10$ and $Z_p=X_p,p>10$.

We observe $Z_1,\dots,Z_{15}$.  Depending on the value of a latent variable, a connection becomes active or `switched on',  if it crosses some cutoff  and  remains `switched off' or inactive otherwise; namely, the connections: $1\leftrightarrow 4$, $1\leftrightarrow 7$, $2\leftrightarrow 5$, $2\leftrightarrow 8$. Here, $i \leftrightarrow j$ or $X_i \leftrightarrow X_j$,  implies that there is an edge between $i$ th and $j$ th node. 
Let $e_n$ be the average number of such undetected connections for $n$ observations.
 Table \ref{extreme} shows the average number of $e_n$ based on 100 replications for $n=200,300,500$ for  $\{\tau\}=\{0.3\}$, $\{\tau\}=\{0.5\}$ and $\{\tau\}=\{0.9\}$ for quantile based MCMC. Higher quantile is able to detect these connections and has smaller average $e_n$. Also, $e_n$ decreases with $n$,  as with large $n$ small signal is more likely to be detected. We use standardized version of the observations, $t=1$ and $N(0,1)$ for the prior for the coefficients for MCMC and  we use 9000 samples with 5000 burn ins for this particular simulation setting. 
 
\noindent  {\bf Ex 3.a.ii.} We construct variables $X_1,\dots, X_{16}$  in the following hierarchical manner using moving average type covariance structure. For $x_{1i},x_{2i},\dots,x_{15,i}$, the $i$ th observation for $X_1,\dots, X_{15}$, we assume the following hierarchical model: $x_{1,i}/{r_{i},\dots,x_{10,i}/r_{i}}\sim MVN({\bf 0},\Sigma)$, with $\bf{0}$ is a vector of zeros and  $\Sigma_{kl}=.7^{|k-l|}$ and $\frac{1}{r_{i}}\sim Gamma(3,3)$  and $x_{11,i}\dots,x_{15,i}$ are independent normal variable with mean zero and variance 1, and $X_{16,i}=\log(r_i)$. We have $n=350$ and the data is standardized. 
  
\begin{figure}
\centering
\includegraphics[width=1.4in,height=1.6in]{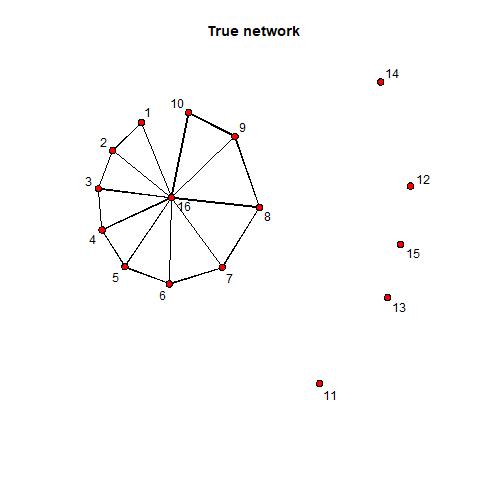}
\includegraphics[width=1.4in,height=1.6in]{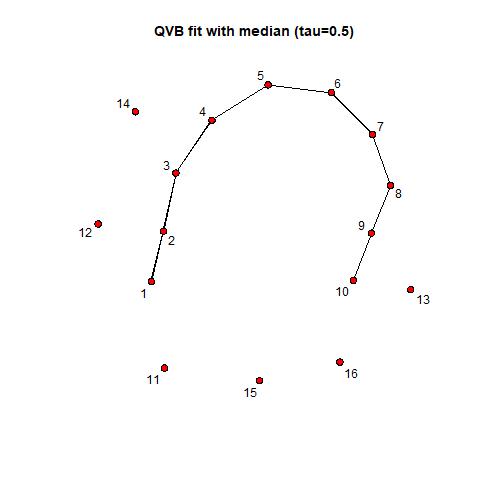}
\includegraphics[width=1.4in,height=1.6in]{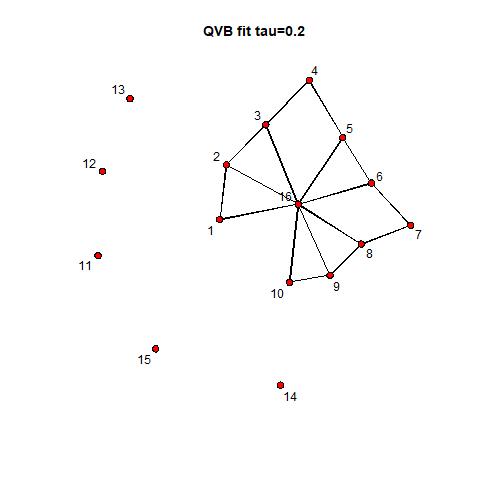}
\includegraphics[width=1.4in,height=1.6in]{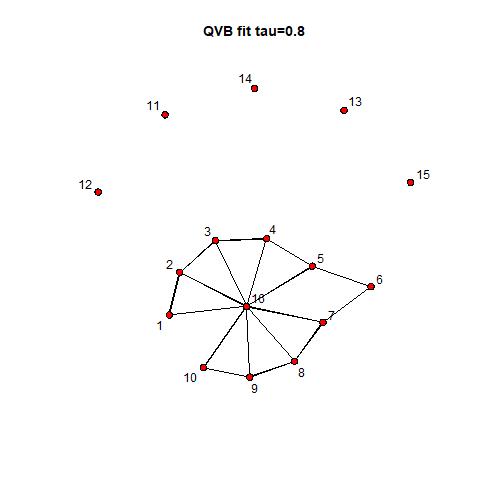}

\caption[Example]{Left panel shows the true network. Second panel shows a sparse fitted graph by variational Bayes for $\tau=\{0.5\}$, and right panel shows the fit using  $\tau=\{0.8\}$  for $n=350$ in Example 3(a)(ii). 
}

\label{qtl_effect}
\end{figure}

  Hence, the network has connections $X_i\leftrightarrow X_j$; $|i-j|<2,i,j\leq10$ and $X_i\leftrightarrow X_{16}$, $i\leq 10$. Using $\tau=\{0.2,0.5,0.8\}$, the QVB fitted network is given in Figure \ref{qtl_effect}. The connections $X_i\leftrightarrow X_{16}$, $i\leq 10$ are not detected for $\tau=0.5$, whereas most of them  are detectable for two extreme quantiles. 

 \begin{figure}
\centering
\includegraphics[width=1.6in,height=1.6in]{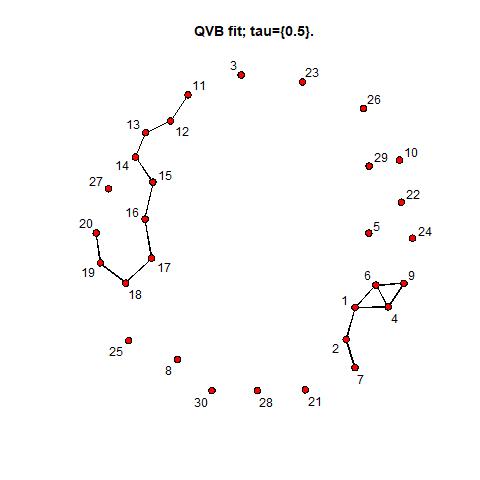}
\includegraphics[width=1.6in,height=1.6in]{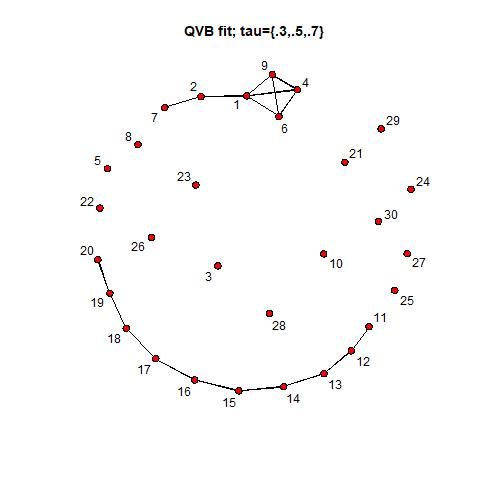}
\includegraphics[width=1.6in,height=1.6in]{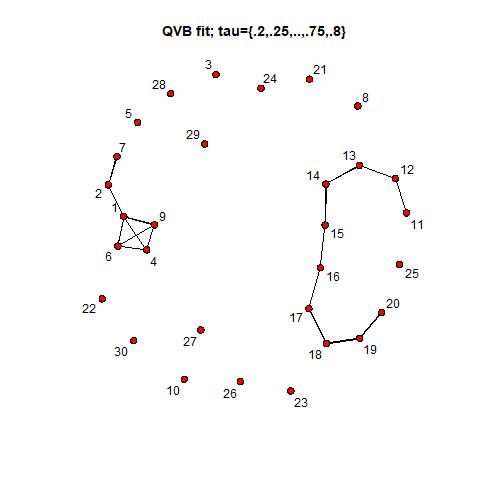}

\caption[Example]{Fit for different quantile grids. Example 3(b)}

\label{granularity}
\end{figure}

%
%
%
%

\subsubsection{\bf Example 3 b.  Granularity of quantile grid}
If we make the quantile grids denser, then we will have neighborhood selected for each of the quantiles and the neighborhood selected would be the union of those neighborhood. But if we use more and more quantiles the FDR stabilizes, as it is implied by Theorem 3.4 and the following Remark \ref{unif_bf}, where we can have the ratio of posterior probability of  any wrong alternate model  with respect to true model, going to zero uniformly over all quantiles, with high probability.  The following examples demonstrate this robustness of quantile-grid selection using the variational Bayes method. 

We consider the set up similar to Example 1(a), with quantile grids of width $0.1$,$.05$ and $.025$, $\{0.2,\dots,0.8\}$ and $\{0.05\}$, $\{0.2,0.25,\dots,0.75,0.8\}$,$\{0.2,0.225,\dots,0.775,0.8\}$. We have $30$ nodes With $X_1,\dots,X_{10}$ generated similar to Ex 1(a), $X_{11}\dots,X_{20}$ follows multivariate normal $MVN(0,\Sigma)$ with $\Sigma_{i,j}=.7^{|i-j|}$, and $X_{21},\dots,X_{30}$ follows independent normal with mean zero and variance one.  A typical QVB fit  for different quantile set up is given in Figure \ref{granularity}, where $\tau=0.5$ captures all but one edge, and $\tau=\{0.3,0.5,0.7\}, \tau=\{0.2,0.25,..,0.75,0.8\}$, gives the correct graph. The FDR's are  given in Table \ref{tabqtl}.   Let $G_1$ be the subgraph based on $X_{1},\dots,X_{9}$, and $G_2$  is the  subgraph based on $X_{11},\dots,X_{20}$,which is disjoint from  $G_1$.  Here $e_1$ is the average number of undetected edges in $G_1$, $e_2$ in $G_2$ and $e_3$ be the average number of connectors detected between them. We can see that the FDR and $e_1,e_2,e_3$,  stabilize even when we increase the number of grids.

\begin{table}
\caption{\label{extreme}A comparison between different quantiles for the extreme value dependence case for Example 3.a using MCMC }
\centering
\begin{tabular}{| c | c    c   c  | }
\hline
$\tau$ & $e_n, n=200$ &$e_n,n=300$ & $e_n,n=500$  \\
\hline
$\{\tau\}=\{.3\}$ &3.88 &3.60 &2.82   \\
$\{\tau\}=\{.5\}$ &2.87 &2.31 &1.22    \\

$\{\tau\}=\{.9\}$&1.65  &0.91  &0.29  \\
\hline

 \end{tabular}
\end{table}

\begin{table}
\caption{\label{tabqtl}Effect of granularity of quantile grid }
\centering
\begin{tabular}{| c | c    c   c   c| }
\hline
Quantile  & FDR & $e_1$ &$e_2$ & $e_3$  \\
\hline
$QVB(\{\tau\}=\{0.5\})$ &0.19 &0.30 &0&0  \\
$QVB(\{\tau\}=\{0.3,.0.5,0.7\})$ &0.41 &0.11 &0&0   \\

$QVB(\{\tau\}=\{.2,.3,\dots,.7,.8\})$ &0.68 &0.10 &0&0   \\

$QVB(\{\tau\}=\{.2,.25,.3,\dots,.7,.75,.8\})$&0.75 &0.10 &0 &0.01  \\

$QVB(\{\tau\}=\{.2,.225,\dots,.775,.8\})$&0.77 &0.10 &0 &0.01 \\

\hline
 \end{tabular}
\end{table}

\subsubsection{Computational gain due to QVB}
In all the cases the variational approximation  based algorithm  performs well to detect the true graphs.  Moreover, QVB is many times faster than the MCMC. We use $40$ iterations for QVB but in all the examples considered, the convergence happens within 20 iterations. Using 5000 samples for each node, and 5000 burn ins, the MCMC runtime is nearly 100 times or more  of  that of the  QVB.   For example for $P=20,60$ and $n=400$ in the set up for example 1, QVB was found to be 170 and 134 times faster over a typical run using one quantile grid. Also, computational cost scales linearly with the number of quantile grids.  Our computation is parallelizable over nodes and the grids of quantiles, though  we do not implement it here. 
%
%
%
%
%
%

\section{Protein network}

The Cancer Genome Atlas (TCGA) is  a source of  molecular profiles for many different tumor types.  Functional protein analysis by  reverse-phase protein arrays (RPPA) is included in TCGA and looking at the proteomic characterization the signaling network can be established.

Proteomic data generated by RPPA across $>8000$ patient tumors obtained from TCGA includes many different  cancer types. We consider lung squamous cell carcinoma (LUSC) data set. The data set considered,  has $n=121$ observations with $P=174$ high-quality antibodies.  The antibodies encompass major functional and signaling pathways relevant to human cancer and a relevant network gives us their interconnection subject to LUSC.  A comprehensive analysis of similar network can be found in Akbani et al. (2014)  for various cancers,  where the EGFR family along with MAPK and MEK lineage was found to be dominant determinant of signaling, where for LUSC it was mainly EGFR. 

We use our quantile based variational approach with $\{\tau_l\}=\{.1,.2,.3,\dots,.7,.8,.9\}$ and a normal prior on $\underline{ \beta }$ (independent $N(0,1)$). Overall, the QVB graph is robust to this prior variance$(v)$ selection in the range $v=[1,100]$ with very few of the edges/weaker connections may be missing for a relatively higher variance. The data is standardized and we use $t=1$.  The graphical LASSO method cannot select a sparse (using huge.select) network both using criterion {\it `MB'} or {\it `GLASSO'} and using criterions for tuning parameter selection. Choosing the penalization by direct  cross validation in GLASSO   in Akbani et al. (2014),  the network has been generated and it reports the important connections.

\begin{figure}
\centering
\includegraphics[width=6in,height=3.5in]{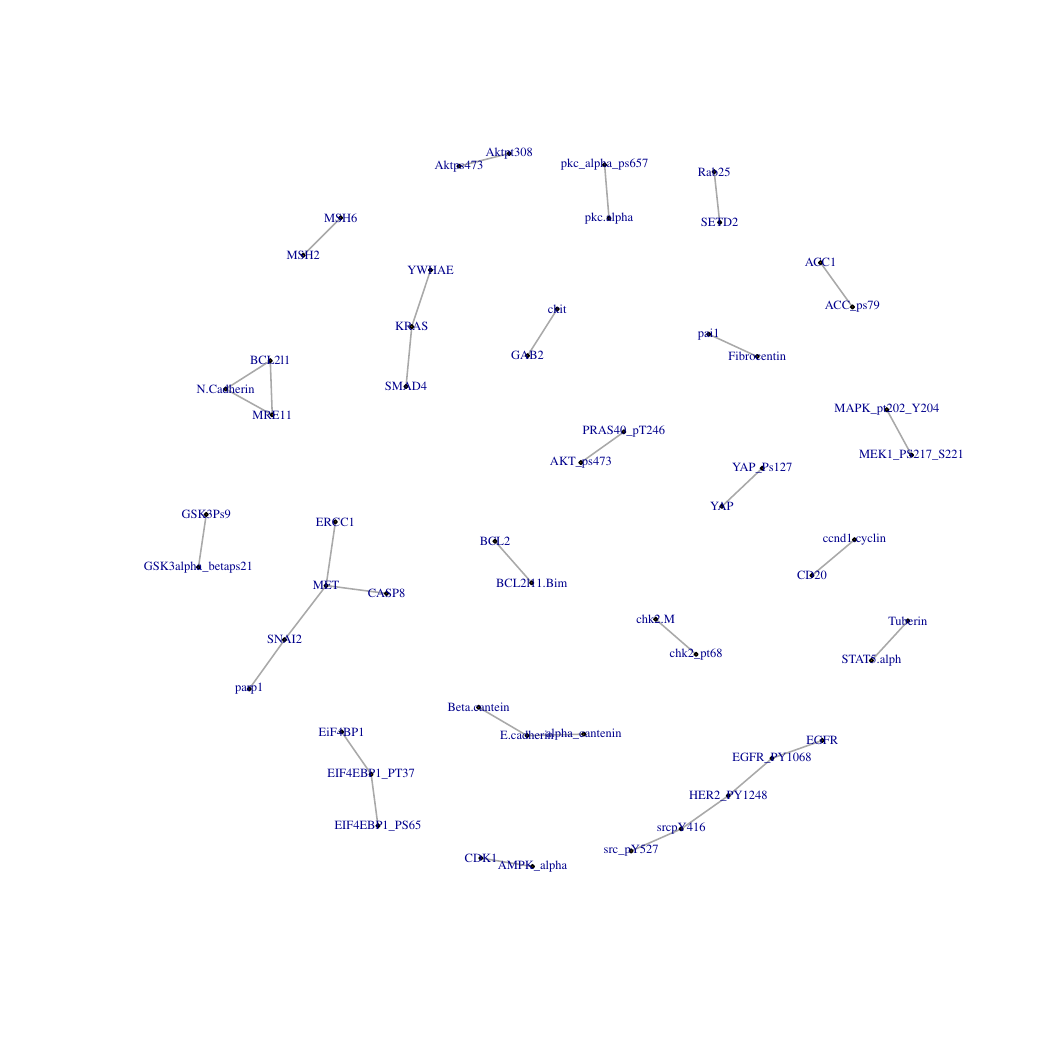}
\caption[Example]{Active proteins and connections for the LUSC data set.}
\label{LUSC}
\end{figure}

 The network and the connection tables with variable index can be found in Figure  \ref{LUSC}  and Table \ref{tab4}. The type of the connection (positive/negative) is also provided.  We can say that one variable effects other variable positively (negatively), conclusively,  if  the coefficients  in  the corresponding quantile regression is greater (less) than  zero for at least one  quantile, and greater (less) than equal to zero for other quantiles. A network of protein was established for different cancer types in  Akbani et al. (2014), where important connections were established. We compare our network for the LUSC network from Akbani et al.  where we find some of the known established connections are detected and also some connection not mentioned  in Akbani et al. have been detected.  Though we refrain from making any inferential claim about the new connections, some further study may be helpful for possibly new biological insight.

In our fitted network, the strong  EGFR/HER2 connections are detected as seen in Akbani et al. (2014). The  connection between $EGFRp$$Y1068$ and     $HER2p$ $Y1248$ is detected which are known to cross react.  
 The connection between $E. Cadherin$ and $alph/beta.Catenin$ is detected as expected. Unlike Akbani et al., $pAkt$ and $Pras40$ are found to be connected in LUSC. This connection was reported for few other cancer types. Also, $MEK$ is  active and connected to $PMAPK$.  $EGFR$ is known to be active in lung cancer and mutation of $MAPK$ , $MEK$ are known to be present for various cancers (see Yatabe et al. (2008), Hilger et al.(2002) ).  Few connections, such as the new negative connection between $p85$ and $claudin7$, mentioned in Akbani et al. (2014) are not detected in this current set up.

  We have detected some new connections not given in Akbani et al. for LUSC data set, such as between $SNAI2$ and  $PARP1$.  Here, SNAI2 is a DNA-transcriptional repressor and $PARP1$ modulates  transcription. 
%

Proteins, $casp8$ and $ERCC1$ are found to be connected with $MET$, which are not given in AKbani et al. $casp8$ performs protein metabolism and  $ERCC1$ is related to structure specific DNA repairing and  known to be important in lung cancer  treatment (Ryu et. al. (2014)). They both are connected to growth factor receptor $MET$.   The detected connections between $KRAS$ and $smad4$, $YWHAE$ and $KRAS$  are not given in the network from Akbani et al. for LUSC data set  and need further study.

\begin{table}
\caption{\label{tab4}Connections and corresponding nodes in LUSC data set}
\centering
\begin{tabular}{|   c |c| }
\hline

 Proteins &Sign      \\

\hline

  $MSH6  \leftrightarrow  MSH2$ &  +\\
\hline
  $AktPT308  \leftrightarrow AktPS473  $ & + \\
  \hline

     $ACC1 \leftrightarrow ACC$&+   \\
    \hline
   $beta.cantenin \leftrightarrow E.Cadherin $& + \\
    
\hline

    $ SNAI2 \leftrightarrow PARP1$&+ \\
 \hline

   $CCND1.Cyclin \leftrightarrow CD20 $ & + \\
    
\hline

   $SrcpY416 \leftrightarrow Src $ & + \\

\hline

     $GSK3.alpha.beta.pS21 \leftrightarrow GSk3pS9$ & + \\
\hline
  $Pai1 \leftrightarrow Fibrocentin $ & +  \\
\hline

     $PRAS40\_ pT246 \leftrightarrow Akt\_pS473$ & + \\
\hline


      $Chk2\_pT68 \leftrightarrow chk2.M$ &+ \\
\hline
$Tuberin \leftrightarrow STAT5.alpha$ & +\\

%
\hline

       $YAP\_PS127 \leftrightarrow YAP$ & +\\

\hline

    $ AMPK\_alpha \leftrightarrow CDK1 $ & + \\
\hline

    $AMPK\_PT172 \leftrightarrow AMPK\_alph $ & + \\
\hline

 $ EIF4EBP1\_pT37  \leftrightarrow  EIF4EBP1 $ & +\\
\hline

 $ EIF4EBP1\_pT37  \leftrightarrow  EIF4EBP1\_PS65 $ & +\\
\hline
 $ alpha.Catenin  \leftrightarrow  E.Cadherin $ & +\\
\hline
 $ cKit  \leftrightarrow  GAB2 $ & +\\
\hline

  $HER2\_pY1248  \leftrightarrow  EGFR\_PY1068$ &  +\\
    
\hline
   $HER2\_pY1248  \leftrightarrow  Src\_PY416$ &  +\\
    
\hline
 
  $MET  \leftrightarrow  SNAI2$ &  +\\
    
\hline
  $MET  \leftrightarrow  CASP8$ &  +\\
    
\hline
 $ERCC1  \leftrightarrow  MET$ &  +\\
%
%
\hline
 $BCL2  \leftrightarrow  Bim$ &  +\\
    \hline
 $KRAS  \leftrightarrow  YWHAE$ &  +\\
    
\hline

 $KRAS  \leftrightarrow  Smad4$ &  +\\

\hline
%
%
    
  $MAPK\_pT202\_Y204 \leftrightarrow MEK1\_pS217\_S221$ &  +\\
    
\hline
 $PKC.alpha\_pS657 \leftrightarrow PKC.alpha$ &  +\\
   
\hline
  $EGFR \leftrightarrow EGFR_pY1068$ &  +\\

%
\hline
  $Rab25 \leftrightarrow SETD2$ &  +\\
    
%
\hline 
$N.Cadherin \leftrightarrow BCL2$ & +\\
\hline
$N.Cadherin \leftrightarrow MRE11$ & +\\
\hline

\end{tabular}

\label{LUSCt}

\end{table}

\clearpage

\section{Discussion}
The proposed approach offers a robust, non-Gaussian model as well as easily implementable algorithms for  sparse graphical models. Even with  large values of  $P$ and relatively smaller value of $n$, it is possible to detect underlying connections as shown in example 1 and in the analysis of the LUSC data set. In the protein network construction, we are able to establish the known signaling network with some newly discovered connections, which need to be validated. 

In this development, we prove the  density  estimation and neighborhood selection  consistency and posterior concentration rate  under both the true model and the misspecified model.  Under misspecified model the posterior concentration  occurs around the  minimum KL distance point from the true density and the set of proposed densities. From simulation examples where we do not assume any density structure in the data generating model, the proposed method performs well. In future, we will further investigate the model selection properties for each node and related convergence rate. 
%

\newpage
\section{Appendix}
\section*{Proof of the theoretical results}
\subsection*{\bf Proof of Theorem 3.1:}

The sketch of the proof is following. At first we construct  the KL neighborhood and show that it has  sufficient prior probability. The sieve is constructed thereon and outside the sieve the prior probability is decreased exponentially. The construction from  Jiang (2007) can be used as long as an equivalent KL ball around the true density can be constructed under the quantile model. 

First, we show  our calculation for the neighborhood construction of $k$ th node. 
Let $h^*_l=\bx'{\underline{\bf \beta}^*}^{(-k)}_l$, $h_{1,l}^*=\bx'{\underline{\bf \beta}^*}_{\gamma_n,l}^{(-k)}$,$h_{2,l}^*=\bx'{\underline{\bf \beta}^*}_{\gamma_n^c,l}^{(-k)}$.  Here, coefficient vector with subscript $\gamma_n$ denotes the coefficient is set to be zero if the corresponding variable is not in $\gamma_n$. Similarly, it is defined for the subscript $\gamma_n^c$. Also, $\bx$ denotes a generic row of ${\bf X}={\bf X_{-k}}^*$.

In model $\gamma_n$, let  $H$ be the set of $\underline{\bf \beta}^{(-k)}_l$'s such that $\beta_{j,l}^{(-k)}\in ({\beta_{j,l}^*}^{(-k)}\pm\eta\frac{ \epsilon_n^2}{r_n})$ for $j \neq k \in \gamma_n$, where $|\gamma_n|=r_n$, such that $\Delta_k^l(r_n)$ is minimized.  Let $h_l=\bx'{\underline{\bf \beta}}_{\gamma,l}^{(-k)}$ then for $\underline{\bf \beta}_l^{(-k)}$ in $H$, we have 

\[L(f)=|log  \frac{f(x_k,h^*_l)}{f(x_k,h_l)}|=|log \frac {f(x_k,h^*)}{f(x_k,h_{1,l}^*)}\frac {f(x_k,h_{1,l}^*)}{f(x_k,h_l)}|\]
\[\leq t\Delta^l_k(r_n)\sum_{j \notin \gamma_n}|X_j|+ \sum _{j \in \gamma_n} t\eta \frac{\epsilon_n^2}{r_n}|X_j|,\]
where $f(x_k,h)\propto \tau(1-\tau)exp(-t\rho_\tau(x_k-h))$.
This step follows from the Lemma 1 of Sriram et al. (2013). 

Therefore, \[ \int L(f)  f^*(x_k|-k) f^*(x_{i\neq k})d{\bf x} \leq  tM^*( p_n \Delta^l_k(r_n)+\eta \epsilon_n^2)<\epsilon^2_n\]
for some appropriately chosen $\eta$ (by A7).    Hence, $H$ lies in the $\epsilon^2_n$ KL neighborhood. 

 For normal prior on the coefficient, $\Pi(H)\geq e^{-cn\epsilon_n^2}$
and $\Pi(\gamma=\gamma_n)>exp(-cn\epsilon_n^2)$ for any $c>0$ for large $n$, similar to Jiang(2007). Therefore, they provide sufficient prior mass on small KL neighborhood around the true density. 

Let  $\tilde{P}_n$ be the set such that regression coefficients lies in $[-C_n,C_n]$ and $\bar{r}_n$ is the maximum model size. For $\delta=\eta\frac{ \epsilon_n^2}{\bar{r}_n}$  covering each of the coefficients by $\delta$ radius $l^\infty$ balls, in those balls we have Hellinger distance less than $\epsilon_n^2$. Hence, we have the total Hellinger covering number of $
\tilde{P}_n$ as  $N(\epsilon_n)\leq\sum_{r=0}^{\bar{r}_n}{p_n^{r}}(\frac{2C_n}{2\delta}+1)^{{r}}\leq(\bar{r}_n +1)({p_n}(\frac{C_n}{\delta}+1))^{\bar{r}_n}$
 (see Jiang; 2005, 2007). 

This step follows as $d^2(p,q)\leq KL(p,q)$, where $d$ is the Hellinger metric defined in section 3 and KL is the Kullback-Leibler distance.   Note that, $log(N(\epsilon_n))=\mathcal{O}(\bar{r}_n(log(C_n)+log(p_n)+log(1/\epsilon_n^2)))$.

 We have from A1, A2, using $C_n$ as a large power (greater than one) of $n$, from Jiang (2005), or any $K_1>0$, for large $n$ 
\begin{eqnarray*}
log(N(\epsilon_n),\tilde{P}_n) &\prec& n\epsilon_n^2,\\
\Pi(\tilde
{P}_n^c) &\leq& e^{-K_1 n\epsilon_n^2 }.
\end{eqnarray*}

 Therefore, Theorem 3.1 follows from verification of conditions for Theorems 5, 6 and Proof of Theorem 3, from Jiang (2005) or Proposition 1 from Jiang (2007).
 
 From Proposition 1  part (i) from  Jiang (2007),
  $P^*[\Pi(h_k > \epsilon_n|D_n)>e^{-c_1'n\epsilon_n^2}] \rightarrow 0$ for some $c_1'>0$. 

\subsubsection*{ Beta-Binomial Prior}
We have shown the result for $I_{j,l}\sim Bernouli(\pi_n)$ with $r_n=p_n\pi_n,{r}_n$ satisfying A1--A7. We use the same ${r}_n$ for Beta-Binomial prior calculation. For $\gamma_n$  with model size ${r}_n$, constructed as in proof of Theorem 3.1, we show that the prior mass condition holds. 

For Beta-Binomial prior on $I_{j,l}$, we have for $a_1=b_1=1$,
\[\Pi(\gamma=\gamma_n)\geq\big((p_n+1) \binom{p_n}{{r}_n}\big)^{-1}.\]
From A1,  $\Pi(\gamma=\gamma_n)>(p_n+1)^{-({r}_n+1)}>exp(-cn\epsilon_n^2)$ for any $c>0$ for large $n$. Therefore, the condition on prior mass holds. Hence, from the earlier proof, Theorem 3.1 follows. 

\subsection*{Proof of Theorem 3.2}
We prove this part for fixed $t_l=t$, and without loss of generality $t$ is assumed to be 1. For Theorem 3.2 and 3.3 we first prove under the assumption of bounded covariate with $|X_k|<M$ for a simplified proof. Later, we relax the condition to accommodate sub exponential tail bound. To show the concentration of $f_{l,k,-k}$ under $\Pi_l(\cdot)$, around $f^*_{l,k,-k}$  the closest point in conditional quantile based likelihood for $\tau_l$, we drop the suffix $l$ in $f_{l,k,-k}(\cdot)$, $\underline{\bf \beta}_l$, $\delta^*_{k,l}$ and $\Pi_l(\cdot)$ for convenience and show for one general quantile.

We have 
\begin{eqnarray}
\Pi(KL( f^0_{k,-k}f^0_{-k},f_{k,-k}({\underline{\bf \beta}}^{(-k)} )f^0_{-k})>\delta+\delta_k^*|.)&= &\frac{ \int_{K_\delta^c}f^n_{k,-k}(\underline{\bf \beta}_\gamma^{(-k)})\Pi(\underline{\beta},\gamma)d(\underline{\beta},\gamma)}{\int f^n_{k,-k}(\underline{\bf \beta}_\gamma^{(-k)})\Pi(\underline{\beta},\gamma)d(\underline{\beta},\gamma)} \nonumber\\
&\leq& \frac{ \int_{K_\delta^c}f^n_{k,-k}(\underline{\bf \beta}_\gamma^{(-k)})\Pi(\underline{\beta},\gamma)d(\underline{\beta},\gamma)}{\int_{v_{\delta',\gamma_0}}f^n_{k,-k}(\underline{\bf \beta}_\gamma^{(-k)})\Pi(\underline{\beta},\gamma)d\underline{\beta}}\nonumber \\
&=&\frac{N_n}{D_n}.
\label{ms_pf1}
\end{eqnarray}

Here, $\gamma_0$ the  vector of $0$ and $1$ corresponding to the  active set (i.e present in the model) of covariates for $k$ th node for the model with the KL distance $\delta_k^*$, and let $|\gamma_0|=M_0'$, the cardinality of the active set and $v_{\delta',\gamma_0}$  be the set with $\gamma_0$  active and  where $\|\underline{\bf \beta}^{(-k)}_{\gamma_0}-\underline{\hat{\bf \beta}}^{(-k)}_{\gamma_0}\|_\infty<{\delta'}{}$. 
Here,  ${K_\delta^c}$ denotes the set of densities where the $KL$ distance from the $f^0_{k,-k}f^0_{-k}$ is more than $\delta+\delta_k^*$. On $K_\delta^c$, $E_{f^0}(log\frac {f^*_{k,-k}}{f_{k,-k}})=KL(f^0_{k,-k}f^0_{-k},f_{k,-k}f^0_{-k})-KL(f^0_{k,-k}f^0_{-k},f^*_{k,-k}f^0_{-k})>\delta$.  Also,  $n$ in the density $f^{n}_{k,-k}$ denotes the likelihood based on $n $ observations. We divide $N_n$ and $D_n$ by $f_{k,-k}^{n*}$ which is likelihood based on $n$ observations under this minimum KL distance model at $k$ th node for $\tau_l$.

Under Beta-Binomial prior $\Pi(\gamma=\gamma_n)>\big((p_n+1) \binom{p_n}{M_0'}\big)^{-1}>e^{-M_0log(p_n+1)}.$
Note that if the difference between coefficient vectors is $\delta^*$ in supremum norm, then the difference between corresponding  log likelihood is at most $MM_0\delta^*$, by B1 and lemma 1(b) from Sriram et al (2013), if we assume $M>1$ without loss of generality.  

Therefore, for the denominator, we have $e^{nc\delta'}\frac{D_n}{f^{n*}_{k,-k}}>e^{-M_0log (p_n+1)}\Pi(v_{\delta',\gamma_0})$  $e^{n(c-M(M_0+1))\delta'}>1$ if $c>M(M_0+1)$, for large $n$. 

We split the numerator in two parts. First part contains the part where each of the entry of $\underline{\bf \beta}^{(-k)}$ lies in $[-K_c-m_\beta,K_c+m_\beta]$ a compact set with $m_\beta= \|\underline{\hat{\bf \beta}}^{(-k)}\|_\infty$ and $K_c>0$.   We denote the set by ${\bf S}$ and its compliment by ${\bf S}^c$. Also, let $m^s_\beta=sup_k \|\underline{\hat{\bf \beta}}^{(-k)}\|_\infty$. For notational convenience, we will drop the index $k$ from the coefficient.

\subsubsection*{ Calculation on ${\bf S}$:}

For any of the at most $c_n\leq2^{M_0}{p_n-1 \choose M_0}$ many covariate combinations ( a conservative bound)  for the $k$th node, we show the part in the ${\bf S}$ decreases to zero exponentially fast. Note that ${p_n-1 \choose M_0}<p_n^{M_0}$.  For any covariate combination, we break the $M_0$ dimensional model space in ${\bf S}$ in ${(MM_0)}^{-1}\delta^{''}$ width $M_0$ dimensional squares. 

Let, $J_n(\delta^{''})$ be the number of squares and for density  $f_{k,-k}$ associated with each nodal point of the ${(MM_0)}^{-1}\delta^{''}$ width  grids, we have $\frac{f^n_{k,-k}} {f^{n*}_{k,-k} }\leq e^{-.5n\delta}$ for large $n$ with probability one, as $n^{-1}log\frac{f^n_{k,-k}}{f^{n*}_{k,-k}}\rightarrow E_{f^0}(log\frac {f_{k,-k}}{f^*_{k,-k}})<-\delta$.

 Also, over all possible covariate combinations: $P(\frac{f^n_{k,-k}} {f^{n*}_{k,-k} }> e^{-.5n\delta},$ $\text{ infinitely often (i.o)})$  $=P(n^{-1}log\frac{f^n_{k,-k}}{f^{n*}_{k,-k}}>{-.5\delta}, \text{i.o})\leq P(n^{-1}log\frac{f^n_{k,-k}}{f^{n*}_{k,-k}}-E_{f^0}(log\frac {f_{k,-k}}{f^*_{k,-k}})>{.5\delta}, \text{i.o})\leq 2^{M_0}lim\sum_n^\infty p_n^{M_0}e^{-c\frac{n\delta^2}{t_s^2}}$  $\rightarrow 0$, by Hoeffding inequality and Borel-Cantelli lemma using $B4$. Here, $t_s=M_0(M+1)(K_c+m^s_\beta)$ and $c>0$ is a generic constant  and $|log\frac{f_{k,-k}}{f^{*}_{k,-k}}|<2t_s$.

 For any point  $\underline{\bf \beta}=\underline{\bf \beta}^{(-k)}$ and its nearest  grid point $\underline{\bf \beta}_{grid}$, we have$\frac{f^n_{k,-k}(\underline{\bf \beta})} {f^{n}_{k,-k}(\underline{\beta}_{grid})}\leq e^{n\delta^{''}}$ (an application of Lemma 1(b), Sriram et al.,2013). 
 
 Choosing $\delta^{''}$ less than $.25\delta$, for large $n$ we have  
  for all the combinations of $\gamma'$ on ${\bf S}$, 
$pr_{.,n}=\sum_{\gamma'}$ $\int_{{\bf S},\gamma'}\frac{f^n_{k,-k}} {f^{n*}_{k,-k} }\Pi(.)$  $\leq  e^{M_0log p_n}J_n(\delta^{''})e^{-nd_1\delta}$ where $ d_1>0$ is a constant depending upon $\delta^{''}$ . Also, $log p_n\prec n$. 
Therefore, $pr_{.,n} < e^{-.5nd_1\delta}$ almost surely.
\vspace{0.1in}

\subsubsection*{Calculation on ${\bf S}^c$:}

Next, we look at ${\bf S}^c$.
Let, $c_\tau=\text{ min}_l\{\tau_l,1-\tau_l\}$. On ${\bf S}^c$, at least one $\beta_{i,l}^{}$ is outside $[-K_c-m_\beta,K_c+m_\beta]$.
 Without loss of generality we assume $\beta_{i,l}^{}-\hat{\beta}_{i,l}^{}$'s  have same sign as $\beta_{0,l}^{}-\hat{\beta}_{0,l}^{}$  (otherwise we change $X_{i}'=-X_i$ and work with the reflected variable). Without loss of generality, we denote the covariate $X_{i}', i\neq k$ encompassing both reflected and non reflected scenarios. As there are only finitely many orderings, it is sufficient to consider only one such case and prove in that case.  Furthermore without loss of generality, the variables are assumed to be centered. 

First, we consider the case when $\beta_{0,l}^{}>\hat{\beta}_{0,l}^{}$.  The case  $\beta_{0,l}^{}<\hat{\beta}_{0,l}^{}$ follows identically. 
We show our calculation for  $\tau_l$,  $y=X_k$ and the covariates $X_1',X_2'$, when $M_0=2$. For general $M_0$, it follows similarly.
Let $b_{i,l}=\beta_{0,l}-\hat{\beta}_{0,l}+(\beta_{1,l}-\hat{\beta}_{1,l})x_{1,i}'+(\beta_{2,l}-\hat{\beta}_{2,l})x_{2,i}'$. Note that if $x_{1,i}',x_{2,i}'>\epsilon>0$, then $b_{i,l}>0$. 

Let
\[\delta_{i,l}=\rho_{\tau_l}(y_i-\beta_{0,l}-\beta_{1,l}x_{1,i}'-\beta_{2,l}x_{2,i}')-\rho_{\tau_l}(y_i-\hat{\beta}_{0,l}-\hat{\beta}_{1,l}x_{1,i}'-\hat{\beta}_{2,l}x_{2,i}').\]
Then from  Lemma 1 and Lemma 5 from Sriram et al. (2013)
\begin{eqnarray*}
\delta_{i,l}&\geq&c_\tau \epsilon K_c {I }_{x_{1,i}'>\epsilon, x_{2,i}'>\epsilon}-2|y_i-\hat{\beta}_{0,l}-\hat{\beta}_{1,l}x_{1,i}'-\hat{\beta}_{2,l}x_{2,i}'|.
\end{eqnarray*}
Let, $A^i_\epsilon=\{x_{1,i}'>\epsilon, x_{2,i}'>\epsilon\}$ and $B^i_\epsilon=\{x_{1,i}'<-\epsilon, x_{2,i}'<-\epsilon\} $.  The previous step follows from the proof of the Lemma 1 in Sriram et al. (2013) by  writing down the loss function explicitly and from  the fact that $b_{i,l}=\tilde{\mu}^{(2)}_{i,l}-\tilde{\mu}^{(1)}_{i,l}>0$ on  $A^i_\epsilon$, where $\tilde{\mu}^{(1)}_{i,l}= \hat{\beta}_{0,l}+\hat{\beta}_{1,l}x_{1,i}'+\hat{\beta}_{2,l}x_{2,i}'$ and $\tilde{\mu}^{(2)}_{i,l}=\beta_{0,l}+\beta_{1,l}x_{1,i}'+\beta_{2,l}x_{2,i}'$. Considering the ordering of $y_i, \tilde{\mu}^{(1)}_{i,l},\tilde{\mu}^{(2)}_{i,l}$, such as $y_i\leq\tilde{\mu}^{(1)}_{i,l}\leq\tilde{\mu}^{(2)}_{i,l}; \tilde{\mu}^{(1)}_{i,l}\leq y_i\leq\tilde{\mu}^{(2)}_{i,l}$  and so on, the above claim can be verified.

 Let $\text{min} \{E(I_{A^i_\epsilon}),E(I_{B^i_\epsilon})\}=a_\epsilon>0$ (by $B3$, choosing appropriate $\epsilon>0$) and $r_i=|y_i-\hat{\beta}_{0,l}-\hat{\beta}_{1,l}x_{1,i}'-\hat{\beta}_{2,l}x_{2,i}'|$ and $E(r_{i})\leq\epsilon'$, over all nodes and  all possible model combination  of size at most $M_0$, at each node (follows from uniformly bounded $\|\hat{\underline{\bf \beta}}^{(-k)}\|_\infty$ and uniformly bounded second moments).

\subsubsection*{ Establishing bound on the average of the indicators and $r_i$}

By Hoeffding bound $P(\sum_{i=1}^n n^{-1}(I_{A^i_\epsilon})<\frac{a_\epsilon}{2})<e^{-2n\frac{a_\epsilon^2}4}$ and similar bound holds for $B_\epsilon^i$. 
Similarly, $P(n^{-1}\sum_{i=1}^ n(r_i)>{2\epsilon'})<e^{-c_2n(\epsilon')^2}$ for some constant $c_2$ as $X_k$'s are bounded. Hence by Borel-Cantelli lemma, the probability $ \frac{n^{-1} \sum _{i=1}^n(I_{A^i_\epsilon})}{n^{-1}\sum_{i=1}^n r_i}<\frac{a_\epsilon}{4 \epsilon'} $ infinitely often  is less than \[{lim}_{N\rightarrow \infty} \sum_{n=N}^\infty p_n^{M_0}(e^{-c_2n(\epsilon')^2}+ e^{-2n\frac{a_\epsilon^2}4}) \rightarrow 0\] by B4.  

Therefore for all the possible at most $M_0$ neighbors, we have, $\frac{n^{-1} \sum_{i=1}^n (I_{A^i_\epsilon}) }{n^{-1}\sum_{i=1}^n r_i}>d_0>0$ for some $d_0$ for all but finitely many cases, with probability 1. The calculation holds for each quantile.  

Also, $ \text{lim} \sum_{n=N}^\infty p_np_n^{M_0}(e^{-c_2n(\epsilon')^2}+ e^{-2n\frac{a_\epsilon^2}4}) \rightarrow 0$ similarly. Therefore, this result holds over the union over   all the vertices /nodes of the graph, over all possible model combination of maximum size $M_0-1$.

Hence choosing $K_c$ large enough,  on ${\bf S}^c$ we have   $log f^n_{k,-k}()-log{f^{n*}_{k,-k}}=-\sum_{i} \delta_{i,l} \leq - n u_0$, where $u_0>0$,  for large  $n$, almost surely.

Therefore choosing $\delta'<\frac{min\{u_0,.5d_1\delta\}}{2M(M_0+1)}$, from \eqref{ms_pf1} LHS goes to zero almost surely, as $e^{nc\delta'}\frac{N_n}{f^{n*}_{k,-k}} \rightarrow 0$, by choosing $c=1.5M(M_0+1)$.

\subsection*{Proof of Theorem 3.3}
This proof follows similar construction of ${\bf S}$ and ${\bf S}^c$ from the previous proof of Theorem 3.2. Here we show for bounded $X_k$'s first. 

\noindent \underline{\bf On ${\mathbf S}$}

For any of the at most $c_n\leq 2^{M_0}{p_n-1 \choose M_0}$ many covariate combinations for the $k$th node, we show the part in the ${\bf S}$ decreases to zero exponentially fast.   We break the $M_0$ dimensional model space in ${(MM_0)}^{-1}\delta^{''}$ width $M_0$ dimensional squares. 

Let, $J_n(\delta^{''})$be the number of squares and for each nodal point we show $\frac{f^n_{k,-k}} {f^{n*}_{k,-k} }\leq e^{-2n\epsilon_n^2}$ almost surely.  This step follows from the following application of Hoeffding inequality.  Note that, $J_n(\delta^{''})=\mathcal{O}((\frac{1}{\delta^{''}})^{M_0})$.

\vspace{0.2in} 

\subsubsection* {Showing  $\frac{f^n_{k,-k}} {f^{n*}_{k,-k} }\leq e^{-n\epsilon_n^2}$ for large $n$ on {\bf S}}

Let, $t_m=M_0(M+1)(K_c+m_\beta)$, $t_s=M_0(M+1)(K_c+m^s_\beta)$. We have  $ E(n^{-1} log(\frac{f^n_{k,-k}} {f^{n*}_{k,-k} }))<-4\epsilon_n^2$.

 Then,  $P( n^{-1} log(\frac{f^n_{k,-k}} {f^{n*}_{k,-k} })>-2\epsilon_n^2, \text{ for some  grid points})< c_{M_0}J_n(\delta^{''})e^{-2n\frac{\epsilon_n^4}{4t_m^2}}$. Here, $c_{M_0}$ is the number of grid points associated with a $M_0$ dimensional grid and $|log\frac{f_{k,-k}}{f^{*}_{k,-k}}|<2t_m$.
 
 Choosing $\delta^{''}=\epsilon_n^2$ , we have   $P(\frac{f^n_{k,-k}} {f^{n*}_{k,-k} }> e^{-2n\epsilon_n^2} \text { infinitely often})  \leq   c_{M_0} \text {lim }_{N\rightarrow \infty}  $ $\sum_{n=N}^\infty $ $J_n(\delta^{''})e^{-2n\frac{\epsilon_n^4}{4t_m^2}} $.  Now from B6 and B7 we get $\sum J_n(\delta^{''})e^{-2n\frac{\epsilon_n^4}{4t_m^2}} <\infty$.  Therefore using Borel-Cantelli lemma,  $\frac{f^n_{k,-k}} {f^{n*}_{k,-k} }\leq e^{-2n\epsilon_n^2}$ almost surely.  
 
 Moreover, $ \sum_n p_np_n^{M_0} J_n(\delta^{''})e^{-2n\frac{\epsilon_n^4}{4t_s^2}}<\infty $ (by $B7$). Therefore, this almost surely convergence happens over all possible covariate combinations and over all $p_n$ vertices/nodes of the graph. 

For any point  $\underline{\bf \beta}$ and its nearest grid point $ \underline{\bf \beta}_{grid} $ ,we have$\frac{f^n_{k,-k}(\underline{\bf \beta})} {f^{n}_{k,-k}(\underline{\beta}_{grid})}\leq e^{n\delta^{''}}$.  Therefore on ${\bf S}$, we have $\frac{f^n_{k,-k}(\underline{\bf \beta})} {f^{n *}_{k,-k}}\leq e^{-n\epsilon_n^2}$, almost surely.  

\subsubsection*{Combining the parts}

\noindent \underline { \it Calculation on  {\bf S}:}

Choosing, $\delta^{'} =.5\frac{\epsilon_n^2}{MM_0}$, we have  
$-log \Pi(v_{\delta',\gamma_0}) =\mathcal{O}(log\frac{1}{\epsilon_n^2})$ and  $e^{nc\delta'}\frac{D_n}{f^*_{k,-k}}>e^{-M_0log p_n}$ $\Pi(v_{\delta',\gamma_0})e^{n(c-MM_0)\delta'}>1$ if $c>MM_0$, for large $n$, from $B6, B7. $

Therefore on ${\bf S}$, choosing $c>MM_0$ and $.5{\epsilon_n^2}<c\delta^{'} <.75{\epsilon_n^2}$

\[\frac{ \int_{K_\delta^c\cap {\bf S}}f^n_{k,-k}(\underline{\bf \beta}_\gamma)\pi(\underline{\beta},\gamma)d(\underline{\beta},\gamma)}{\int_{v_{\delta',\gamma_0}}f^n_{k,-k}(\underline{\bf \beta}_\gamma)\pi(\underline{\beta},\gamma)d\underline{\beta}}\leq e^{-.25n\epsilon_n^2}.\]

\noindent{\underline{ On ${\mathbf S}^c$}

On ${\bf S}^c$, the result from Theorem 3.2 holds and  $log f_{k,-k}(.)-log{f^*_{k,-k}}\leq -n u_0$ almost surely for large $n$.

Therefore, with $n,p_n$ going to infinity,  
 $P( \Pi(KL( f^0_{k,-k}f^0_{-k},f_{l,k,-k}f^0_{-k})>\delta_n+\delta_k^*\text  { for some node }|.)$  goes to zero almost surely. 
\vspace{0.04in}

\subsection*{Relaxing  boundedness condition}

\vspace{0.00in}
From B2, using Holder inequality, we have that for any $M_0+1$ dimensional linear combination of absolute values of  $X_i$'s with bounded coefficient (where coefficient of $X_i$'s are bounded by 1), denoting the random variable by generic symbol $W$:  \[ E(e^{\lambda(W-E(W))})\leq e^{.5\lambda^2{\nu^*}^2} \text{ for } |\lambda|<{b}^{-1}\]  for some  global  $b,\nu^*<\infty$, for all possible such combinations. This is the condition for sub-exponential distribution with parameters $(\nu^*,b)$ with ${\nu^*}^2=(M_0+1)\nu^2$.

\vspace{0.02in}

\noindent \underline {\it Showing for linear combinations}

This result follows from the following argument using Holder's inequality,
 \[E\large(e^{\lambda\sum_{i=1}^M\alpha_i(X_i-E(X_i))\large})\leq E\large(e^{|\lambda|\sum_{i=1}^M|\alpha_i(X_i-E(X_i))|}\large)\leq \{e^{.5\lambda^2M^2\nu^2  }\}^{\frac{1}{M}}\leq e^{.5\lambda^2(M_0+1)\nu^2  }.\]
  Then for $w_1,\dots,w_n$ i.i.d $W$ with mean $\overline{W}$, we have $P(|\overline{W}-E(W)|>t')\leq 2e^{-\frac{n^2t'^2}{2(n(\nu^*)^2+nbt')}}$ (Bernstein-type inequality).  Thus, we induce uniform tail bound on the variables/nodes and their linear combinations. 

\subsubsection *{Showing Theorem 3.3 for sub-exponential tail bound}

From the tail bound result for linear combinations 
\begin{eqnarray}
P(n^{-1}\sum_{i=1}^n(|x_{ij_1}|+\dots+|x_{ij_m}|)>K_M \text{ for some } {j_1,\dots,j_{m}} \in \{1,\dots,p_n\}; m\leq M_0-1)\nonumber \\
\leq p_n^{M_0}e^{-c_1n} \vspace{-.12in}
\label{subexavg}
\end{eqnarray}
with some $c_1>0$, $K_M=1.5M_0\text{max}E(|X_k|)$, as $b<\infty$. Hence, $ n^{-1}\sum_{i=1}^n(|x_{ij_1}|+\dots+|x_{ij_m}|)\leq K_M$ for all but finitely many cases, almost surely by Borel-Cantelli lemma, as $\sum p_n^{M_0}e^{-c_1n}<\infty$. 

We can choose $\delta'=\frac{.5}{K_M+1}\epsilon_n^2$ and for $D_n$, on $v_{\delta',\gamma_0}$ we have $n^{-1}|log f^{n*}_{k,-k}-log f^{n}_{k,-k}|\leq\frac{.5}{(K_M+1)}\epsilon_n^2n^{-1}\sum_i\sum_{j\in \gamma_0}|x_{ij}|\leq .5\epsilon_n^2$ as $n$ goes to infinity (using Lemma 1(b), Sriram et al., 2013).  Similarly, for $N_n$, on ${\bf S}$ we choose ${((K_M+1))}^{-1}\delta''$ size grids and the conclusion for bounded case holds.  

On ${\bf S}$ the absolute value of the coefficients are bounded by $K_{max}=K_c+m^s_\beta$. For linear combination with bounded coefficient, we assumed sub-exponential distribution. Same holds for  differences of such functions with bounded intercept terms, similarly (without loss of generality, we bound the absolute value of coefficients and intercept terms by one, to get the global $b$, $\nu^*$ in sub exponential formulation, using Holder's inequality). We assume global constants $b$, $\nu$ , in the sub-exponential condition, slightly abusing the earlier notation. 

Finally, for each of the grid points, 
\begin{eqnarray*}
P( n^{-1} (log\frac{f^n_{k,-k}}{f^{n*}_{k,-k} })>-2\epsilon_n^2)&=&P( n^{-1}\frac{1}{K_{max}} (log{f^n_{k,-k}}-log {f^{n*}_{k,-k} })>-2\frac{\epsilon_n^2}{K_{max}}) \\
&<&e^{-\frac{c_2}{(K_c+m^s_\beta)^2}n\epsilon_n^4}
\end{eqnarray*} for some fixed $c_2>0$ . This step follows using the  sub-exponential property for the quantile loss functions at nodes and their linear combination, as we have shown it for absolute value  the linear combinations of the covariates earlier, as $b,\nu$ are global constants, in the sub-exponential assumption in this case.    On ${\bf S}^c$ the bound on $n^{-1}\sum_{i=1}^nr_i$ follows similarly, as the intercept $\hat{\beta}_{0,l}$ and the coefficients are bounded.  Hence, the proof of Theorem 3.3 holds under  relaxed assumptions.

\subsection*{Proof of Proposition 2.1}
The proof follows trivially from model given in equation 1 from the main manuscript and the linearity of conditional quantile function. 

\subsection*{Proof of Lemma \ref{lem1}}
Follows readily from the fact that under $C_1$, if $X_j$ is not connected to $X_k$ then $j$ is not contained in any $N^*_{l,k}$, and if $X_j\leftrightarrow X_k$, then  from $C2$, $X_j$ is in some $N^*_{l,k}$ if we choose small enough quantile grid width. 
\subsection*{Proof of Theorem \ref{bf1}}
Let, $M^*_{l,k}$ be the  model for $\tau_l$ at $k$ th node induced by $N^*_{l,k}$, and $M^1_k\neq M^*_{l,k} $ be any competing model at node $k$. Let $f^n_{l,k,-k,M_{1}}(\underline{\beta})=f^n_{l,k,-k,\underline{\beta}_{M_{1}}}$ be the likelihood under Equation \ref{qtllikeli}, for  $n$ observations for coefficient $\underline{\beta}=\underline{\beta}_l$, for some model $M_1$, at node $k$ .  Then, 
\begin{eqnarray}
\Pi^n_{l,k}(M^1_k,M_{l,k}^*)\leq c\pi_n^{-s^*+s_{M_1}}\frac{\int f^n_{l,k,-k,M^1_{k}} (\underline{\beta}) \pi(\underline{\beta})d\underline{\beta} }{ \int f^n_{l,k,-k,M^*_{l,k}} (\underline{\beta})\pi(\underline{\beta})d\underline{\beta}}= c\pi_n^{-s^*+s_{M_1}}\frac{\int \frac{f^n_{l,k,-k,M^1_{k}} (\underline{\beta})}{f^n_{k,-k,{\underline{\hat{\beta}}}_{M^*_{l,k}}}}  \pi(\underline{\beta})d\underline{\beta} }{ \int \frac { f^n_{l,k,-k,M^*_{l,k}} (\underline{\beta}) \pi(\underline{\beta})}{f^n_{k,-k,{\underline{\hat{\beta}}}_{M^*_{l,k}}}}d\underline{\beta}}\nonumber \\
=c\pi_n^{-s^*+s_{M_1}}\frac{N^n_{BF}}{D^n_{BF}}.
\label{eqbf1}
\end{eqnarray}
 Here the suffix $l$ denote the likelihood used corresponds to $\tau_l$, $c<2^{M_0}$ is a constant as $\pi_n\leq0.5$ without loss of generality.  Let  $\hat{\underline{\beta}}_{M_{l,k}^*}$ or the vector of $\hat{\beta}_0,\hat{\beta}_1, \dots, \hat{\beta}_{m^*}$ be the true values of coefficients that minimizes the expected quantile loss and the KL distance with the data generating density. Without loss of generality we can choose them to be first $m^*$ variables. Let $s^*=m^*$ be the size of true model for $k$ th node and $s_{M_1}=s_{M^1_k}$ be the size of the competing model. For convenience we drop the $l$ and writing $M^*_k$ instead of $M^*_{l,k}$, we write $\hat{\underline{\beta}}_{M_{k}^*}$.

For $\Omega^\beta_{\epsilon_n'}=\{\underline{\beta }:\beta_i \in (\hat{\beta}_i\pm {\epsilon_n'}^2/(2K_M)); i=0,\dots,m^*\}$, we have $n^{-1}|log \frac{f^n_{k,-k,\hat{\underline{\beta}}}}{f^n_{k,-k,{\underline{\beta}}}}|\leq {\epsilon'_n}^2$ with probability one, for $\underline{\beta} \in \Omega^\beta_{\epsilon_n'}$ using the fact that for   $K_M=1.5M_0\text{max}E(|X_k|)$, $ n^{-1}\sum_{i=1}^n(|x_{ij_1}|+\dots+|x_{ij_m}|)\leq K_M$ for large $n$ with probability one, following the conclusion following \ref{subexavg}; $m<M_0$.  For bounded covariate,  we  can use $K_M=M_0\text{max}_j|X_j|$. Note that $E|X_j|$ is bounded by $C4$.

As $-log (\Pi(\Omega^\beta_{\epsilon_n'}))=\mathcal{O}(log n)$, for $\epsilon_n'\rightarrow 0$, ${\epsilon'_n}^2 \sim n^{-1+\delta_1}$, $\delta_1>0$ we have $e^{2n{\epsilon'_n}^2}D^n_{BF}>1$ with probability one.  

Next we consider two cases, $M^*_k\subset M_k^1$ and $M^*_k\not \subset M_k^1$.

\subsubsection* {The case {\bf $M^*_k\not \subset M^1_k$}}

Let $\underline{\beta}^n_{M^1_k}$ be the Maximum likelihood estimate of $\hat{\underline{\beta}}_{M^1_k}$, the minimizer of the expected loss under misspecified model.  Then $\underline{\beta}^n_{M^1_k}$ converges to $\hat{\underline{\beta}}_{M^1_k}$ in probability. Consequently,  we show that,  $n^{-1}log \frac{f^n_{k,-k,\underline{\beta}^n_{M^1_k}}}{f^n_{k,-k,\hat{\underline{\beta}}_{M^*_k}}}\leq -\delta$ in  probability,  for some $\delta>0$.

This step follows form the following argument  writing $log \frac{f^n_{k,-k,\underline{\beta}^n_{M^1_k}}}{f^n_{k,-k,\hat{\underline{\beta}}_{M^*_k}}}=log \frac{f^n_{k,-k,\underline{\beta}^n_{M^1_k}}}{f^n_{k,-k,{\underline{\hat{\beta}}}_{M^1_k}}}+log \frac{f^n_{k,-k,{\underline{\hat{\beta}}}_{M^1_k}}}{f^n_{k,-k,{\underline{\hat{\beta}}}_{M^*_k}}}$. Now, $n^{-1}log \frac{f^n_{k,-k,{\underline{\hat{\beta}}}_{M^1_k}}}{f^n_{k,-k,\hat{\underline{\beta}}_{M^*_k}}}<-\delta$ almost surely and hence, in probability, where $l_{\tau,\underline{\hat{ \beta}}_{M^1_k}}-l_{\tau,\underline{\hat{ \beta}}_{M^*_k}}>2\delta$. 

Note that $|log \frac{f^n_{k,-k,\underline{\beta}^n_{M^1_k}}}{f^n_{k,-k,\hat{\underline{\beta}}_{M^1_k}}}|\leq\sqrt{n}n^{-1}\sum_{i=1}^n \sum_{j=1}^{s_{M^1_k}}w_{j,n}|x_{j,i}|$, where $x_{0,i}=1$, where $n^{-.5}w_{j,n}=|\beta^n_{j,M^1_k}-\hat{\beta}_{j,M^1_k}|$. Therefore $n^{.5}n^{-1}|log \frac{f^n_{k,-k,\underline{\beta}^n_{M^1_k}}}{f^n_{k,-k,\hat{\underline{\beta}}_{M^1_k}}}|=\mathcal{O}_p(1)$(Theorem 3, Angrist et. al., 2006; $w_{j,n}=\mathcal{O}_p(1))$, and as a result $n^{-1}log \frac{f^n_{k,-k,\underline{\beta}^n_{M_k^1}}}{f^n_{k,-k,\hat{\underline{\beta}}_{M^*_k}}}\leq -\delta$ in  probability. 

Hence, from equation \eqref{eqbf1}, multiplying numerator and denominator by $ {e^{2n{\epsilon'_n}^2}}$ 

\begin{eqnarray}
\Pi^n_{l,k}(M^1_k,M_{l,k}^*) \leq c\pi_n^{-s^*+s_{M_1}} e^{2n{\epsilon'_n}^2} e^{n[ n^{-1}log \frac{f^n_{k,-k,\underline{\beta}^n_{M^1_k}}}{f^n_{k,-k,\hat{\underline{\beta}}_{M^*_k}}}]}.
\label{eqbf2}
\end{eqnarray}

Hence, $log \Pi^n_{l,k}(M^1_k,M_{l,k}^*)<-\delta'n$ for any $0<\delta'<\delta$ for large $n$ in probability and therefore, $\Pi^n_{l,k}(M^1_k,M_{l,k}^*)$ converges to zero in probability.

\subsubsection* {The case {\bf $M^*_k \subset M_k^1$}}

Without loss of generality assume that $M^1_k$ has first $s_{M_1}>s^*$ variable active. 
Note that, $|log \frac{f^n_{k,-k,\underline{\beta}^n_{M^1_k}}}{f^n_{k,-k,\hat{\underline{\beta}}_{M^*_k}}}|\leq\sqrt{n}n^{-1}\sum_{i=1}^n \sum_{j=1}^{s_{M^1_k}}w_j^n|x_{j,i}|$, where $x_{0,i}=1$, where $n^{-.5}w_{j,n}=|\beta^n_j-\hat{\beta}_{j,M^*}|$. Note that $\hat{\underline{\beta}}_{M^*_k}=\hat{\underline{\beta}}_{M^1_k}$ by uniqueness of the minimizer of expected quantile loss.  

As $w_{j,n}$'s are $\mathcal{O}_p(1)$, therefore, $ n^{-1}\sum_{i=1}^n \sum_{j=1}^{s_{M^1_k}}w_j^n|x_{j,i}|$ is $\mathcal{O}_p(1)$. Hence, from equation \eqref{eqbf2} using B8, 
\[log \Pi^n_{l,k}(M^1_k,M_{l,k}^*)\leq-(s_{M_1}-s^*)c_0n^{.5+\epsilon'}+\sqrt{n}\mathcal{O}_p(1)+2n{\epsilon'_n}^2+c'_0\] for  generic constants $c_0>0,c'_0$.  Choosing $\epsilon'_n<n^{-.25}$, we have
 $\Pi^n_{l,k}(M^1_k,M_{l,k}^*)$  goes to zero in probability.

\subsection*{Proof of Remark \ref{unif_bf}}
Let,   $w_{j,n}(\tau)$ is defined similar to $w_{j,n}$ when we use $\tau$ as our quantile. Note that  for $M_k^*\subset M_k^1$ $sup_{\tau\in (\epsilon,1-\epsilon)}|w_{j,n}(\tau)|$ is $\mathcal{O}_p(1)$, for $\epsilon>0$ and therefore, $\sup_\tau \Pi_{\tau,k}^n(M^1_k,M_{k}^*)$ is $o_p(1)$ from the earlier calculation.  This step follows from the conclusion about the process over $\tau$ in Theorem 3 of Angrist et. al. (2006).

Suppose, we have  minimizer of the quantile loss at $\tau$, $\hat{\underline{\beta}}_{M^*_k}(\tau)$ and $\hat{\underline{\beta}}_{M^1_k}(\tau)$, under $M^*_k$ and $M_k^1$, respectively,  for the case where $M_k^1$ does not contain $M_k^*$. Then, for any $\delta>0$, there exists $\epsilon_1>0$ such that,  $|n^{-1}log \frac{f^n_{k,-k,{\underline{\hat{\beta}}}_{M^1_k}(\tau')}}{f^n_{k,-k,{\hat{\underline{\beta}}}_{M^*_k}(\tau')}}-n^{-1}log \frac{f^n_{k,-k,{\underline{\hat{\beta}}}_{M^1_k}(\tau'')}}{f^n_{k,-k,{\hat{\underline{\beta}}}_{M^*_k}(\tau'')}}|<\delta/4$ for $|\tau'-\tau''|<\epsilon_1$, $\epsilon_1>0$ a small number. This step follows using $n^{-1}\sum_i\sum_{j_l}\sum |x_{j_l,i}|<K_M$ for large $n$ (shown in the proof of Theorem \ref{mis_rate}) and the continuity of  $\hat{\underline{\beta}}_{M^*_k}(\tau)$ and $\hat{\underline{\beta}}_{M^1_k}(\tau)$.

Again,
$n^{-.5}sup_\tau |log \frac{f^n_{k,-k,\underline{\beta}^n_{M^1_k}}}{f^n_{k,-k,\hat{\underline{\beta}}_{M^1_k}}}|\leq sup_\tau  n^{-1}\sum_{i=1}^n \sum_{j=1}^{s_{M^1_k}}|w_{j,n}||x_{j,i}|=\mathcal{O}_p(1)$. Hence, using finitely many $\epsilon_1$ equi-spaced grid at different $\tau$'s  in the set $S_\tau$, for $\tau \in S_\tau$ we have $\Pi^n_{\tau,k}(M^1_k,M_{k}^*)$ goes to zero in probability from the calculation before equation \ref{eqbf2}, by showing $n^{-1}log \frac{f^n_{k,-k,\underline{\beta}^n_{M^1_k}(\tau)}}{f^n_{k,-k,\hat{\underline{\beta}}_{M^*_k}(\tau)}}<-\delta/2$ in probability, for  $\tau$'s  in the set $S_\tau$ for some $\delta>0$. Here, we use the fact $\text{inf}_{\tau\in(\epsilon,1-\epsilon)}l_{\tau,\underline{\hat{ \beta}}_{M^1_k}}-l_{\tau,\underline{\hat{ \beta}}_{M^*_k}}>\delta$ for some $\delta>0$ for a given $\epsilon>0$. Then, we repeat the argument for  the case $M^*_{l,k}\not \subset M^1_k$ in Theorem \ref{bf1} proof. As $S_\tau$ is  a finite set, this convergence to zero  in probability, is uniformly over $S_\tau$.

For $\tau \in (\epsilon,1-\epsilon)\cap S^c_\tau$, from the earlier calculation, 
\begin{eqnarray*}
\Pi^n_{\tau,k}(M^1_k,M_{k}^*)& \leq& c\pi_n^{-s^*+s_{M_1}} e^{2n{\epsilon'_n}^2} e^{n[ n^{-1}log \frac{f^n_{k,-k,\underline{\beta}^n_{M^1_k}(\tau)}}{f^n_{k,-k,\hat{\underline{\beta}}_{M^*_k}(\tau)}}]}\\
&\leq&c\pi_n^{-s^*+s_{M_1}} e^{2n{\epsilon'_n}^2} e^{n[ n^{-1}log \frac{f^n_{k,-k,\underline{\beta}^n_{M^1_k}(\tau)}}{f^n_{k,-k,\hat{\underline{\beta}}_{M^1_k}(\tau)}}+n^{-1}log \frac{f^n_{k,-k,\underline{\hat{\beta}}_{M^1_k}(\tau')}}{f^n_{k,-k,\hat{\underline{\beta}}_{M_k^*}(\tau')}}+\delta/4]} 
\end{eqnarray*}
for $|\tau-\tau'|<\epsilon_1$, $\tau'\in S_\tau$. 
  We have, $sup_{\tau}|n^{.5}n^{-1}log \frac{f^n_{k,-k,\underline{\beta}^n_{M^1_k}(\tau)}}{f^n_{k,-k,\hat{\underline{\beta}}_{M^1_k}(\tau)}}|$ $=\mathcal{O}_p(1)$ and  $n^{-1}log \frac{f^n_{k,-k,\underline{\hat{\beta}}_{M^1_k}(\tau')}}{f^n_{k,-k,\hat{\underline{\beta}}_{M_k^*}(\tau')}}<-\delta/2$ in probability. Hence,  $\Pi^n_{\tau,k}(M^1_k,M_{k}^*)<e^{-n\delta/8}\rightarrow 0$
in probability uniformly over $\tau$. 

\section*{Sequential updates for variational formulation}

For the formulation in Equation \ref{vrupdate} from the main manuscript, we have 
\begin{eqnarray*}
q^*(\underline{\bf \beta}_l)  &\propto& \exp\Big(- \frac{1}{2}E\big((\sum_{i}(y_ i-\bfx_i'\underline{\bf\beta}_{\gamma,l}-\xi_{1,l}v_{i,l})^2)(\frac{t}{v_{i,l}\xi_{2,l}^2})\big)-  \frac{1}{2}\underline{\bf \beta}_l'S_{\beta_l}\underline{\bf \beta}_l)\Big)\\
&=& \exp\Big(-\frac{1}{2}E(\sum_{i}(y_ i-\bfx_{i,\gamma}'\underline{\bf\beta}_l-\xi_{1,l}{E(v_{i,l}^{-1})}^{-1})^2)E(\frac{t}{v_{i,l}\xi_{2,l}^2})) - \frac{1}{2}\underline{\bf \beta}_l'S_{\beta_l}\underline{\bf \beta}_l+c_0\Big)\\
&=& \exp\Big( - \frac{1}{2}E\{({Y^{\delta',l}}-{\bf X_{\gamma}}\underline{{\bf \beta}}_l)' \Sigma_l  ({Y^{\delta',l}}-{\bf X_{\gamma}}\underline{\bf \beta}_l) +\underline{{\bf \beta}}_l'S_{\beta_l}\underline{{\bf \beta}}_l\}+c_0\Big)\\
&=&\exp\Big(- \frac{1}{2}\{(\underline{{\bf \beta}}'E({\bf X_{\gamma}}\Sigma_l{\bf X_{\gamma}})\underline{{\bf \beta}}+\underline{{\bf \beta}}_l'S_{\beta_l}\underline{{\bf \beta}}_l-2\underline{{\bf \beta}}_l'E({\bf X}_{\gamma})'\Sigma_l{Y^{\delta,l}} \}+c_1\Big)\\
&=&\exp\Big( - \frac{1}{2}  \{(\underline{{\bf \beta}}_l-(S_{x,\gamma,l}^E+S_{\beta_l})^{-1}{{\bf X}_{\gamma}^E}'\Sigma_l{Y^{\delta,l}})'(S_{x,\gamma}^E+S_{\beta_l})\\
 & &\text{ \hspace {1.5in }      }(\underline{{\bf \beta}}_l-(S_{x,\gamma,l}^E+S_{\beta_l})^{-1}{{\bf X}_{\gamma}^E}'\Sigma_l{Y^{\delta,l}})+c_2\Big)
\end{eqnarray*}
 where $c_0,c_1$ and $c_2$ are free of $\underline{{\bf \beta}}_l$. Therefore, we have the multivariate normal  form for $\underline{{\bf \beta}}_l$ and hence the result follows.
 
\noindent{\bf For $\pi_l$:}
\begin{eqnarray*}
log(q^*(\pi_l)) =  (a_1+\sum_{j=1,j\neq k}^{P} E( I_{j,l}))log \pi_l +(P-1-\sum_{j=1,j \neq k}^{P}E( I_{j,l})+b_1)) log(1-\pi_l)+c
\end{eqnarray*}
for  some constant $c$ free of $\pi$. Therefore, 
\[q^{new}( \pi_l ) := Beta(a_1+\sum_{j=1,j\neq k}^{P} E( I_{j,l}),P-1-\sum_{j=1,j \neq k}^{P}E( I_{j,l})+b_1).\]
%
%
{\bf For $v_{i,l}$:}

  From  equation 10 from the main manuscript
\[\log q^{*}(v_{i,l})=-\frac{1}{2}\{E(\frac{(y_ i-\bfx_i'\underline{\bf\beta}_{\gamma, l})^2}{\xi_{2,l}^2}){v_{i,l}}^{-1}+E(t_l)(\frac{\xi_{1,l}^2}{\xi_{2,l}^2}+2)v_{i,l}\}-\frac{1}{2}\log v_{i,l} +c',\]
where $c'$ is free of $v_{i,j}$.

Note that inverse Gaussian density with parameter $\mu$ and $\lambda$ has the form
\[f(x,\mu,\lambda) \propto x^{-\frac{3}{2}}\exp(-\lambda\frac{(x-\mu)^2}{2\mu^2x} ){\bf I}_{x>0}\].

Equating the coefficients of $x$ and $\frac{1}{x}$, i.e $v_{i,l}$ and $\frac{1}{v_{i,l}}$, we have 
$\lambda=\lambda_{i,l}=E(t_l)E(\frac{(y_ i-\bfx_i'\underline{\bf\beta}_{\gamma,l})^2}{\xi_{2,l}^2})$ and $\mu=\mu_{i,l}=\sqrt{\frac{\lambda_{i,l}}{2E(t_l)+E(t_l)\frac{\xi_{1,l}^2}{\xi_{2,l}^2}}}.$

\vspace{1in}
\noindent{\bf Indicator function :}

We have,
\begin{eqnarray*}
\log (P(I_{j,l}=1))& =& E(I_{j,l}\log(\pi_l)+(1-I_{j,l}) \log(1-\pi_l))-\\
& & \text{\hspace{.8in}}\frac{1}{2}\{\sum_{{i},I_{j,l}=1}E(t_l)E(\frac{(y_ i-\bfx_i'\underline{\bf\beta}_{\gamma,l}-\xi_{1,l}v_{i,l})^2}{v_{i,l}\xi_{2,l}^2})+c_4\\
&=&E(log\frac{\pi_l}{1-\pi_l})-\frac{1}{2}\{\sum_{{i},I_{j,l}=1}E(t_l)E(\frac{(y_ i-\bfx_i'\underline{\bf\beta}_{\gamma,l}-\xi_{1,l}v_{i,l})^2}{v_{i,l}\xi_{2,l}^2})+{c_4}'
\end{eqnarray*}
where ${c_4}'$ is a constant and
\begin{eqnarray*}
\log (P(I_{j,l}=0)) &=&E(log(1-\pi_l))-\frac{1}{2}\{\sum_{{i},I_{j,l}=0}E(t_l)E(\frac{(y_ i-\bfx_i'\underline{\bf\beta}_{\gamma,l}-\xi_{1,l}v_{i,l})^2}{v_{i,l}\xi_{2,l}^2})+{c_4}\\
&=&-\frac{1}{2}\{\sum_{{i},I_{j,l}=0}E(t_l)E(\frac{(y_ i-\bfx_i'\underline{\bf\beta}_{\gamma,l}-\xi_{1,l}v_{i,l})^2}{v_{i,l}\xi_{2,l}^2})+{c_4}'.\end{eqnarray*}
Therefore, \[ \log(\frac{ P(I_{j,l}=1)}{ P(I_{j,l}=0)}) = E(\log\frac{\pi_l}{1-\pi_l}) -\frac{1}{2}\{\sum_{{i},I_{j,l}=1}E(t_l)E(\frac{(y_ i-\bfx_i'\underline{\bf\beta}_{\gamma,l}-\xi_{1,l}v_{i,l})^2}{v_{i,l}\xi_{2,l}^2}) - \] \[\sum_{{i},I_{j,l}=0}E(t_l)E(\frac{(y_ i-\bfx_i'\underline{\bf\beta}_{\gamma,l}-\xi_{1,l}v_{i,l})^2}{v_{i,l}\xi_{2,l}^2})\}.\]

\setlength{\parskip}{.5 em}

\section*{References}
\begin{enumerate}

\item[] Akbani, R., Ng, P. K. S., Werner, H. M., Shahmoradgoli, M., Zhang, F., Ju, Z., ... , \& Ling, S. (2014). A pan-cancer proteomic perspective on The Cancer Genome Atlas. Nature communications, 5.
\vspace{0.1in}

\item[]Angrist, J., Chernozhukov, V., \& Fern�ndez?Val, I. (2006). Quantile regression under misspecification, with an application to the US wage structure. Econometrica, 74(2), 539-563.
\vspace{0.1in}

\item[] Atay-Kayis, A., \& Massam, H. (2005). A Monte-Carlo Method for Computing the Marginal Likelihood in Nondecomposable Gaussian Graphical Models. Biometrika, 92:317--335.
\vspace{0.1in}

\item[] Barnard, J., McCulloch, R., \& Meng, X. L. (2000). Modeling covariance matrices in terms of standard deviations and correlations, with application to shrinkage. Statistica Sinica, 10(4), 1281-1312.
\vspace{0.1in}

\item[] Beal, M. J. (2003). Variational algorithms for approximate Bayesian inference. Ph.D. thesis, Gatsby Computational Neuroscience Unit, University College London.
\vspace{0.1in}

\item[] Bernardo, J. M. (1979). Expected information as expected utility. The Annals of Statistics, 686-690.
\vspace{0.1in}

\item[] Brooks, S.P., Giudici, P., \& Roberts, G.O.(2003). Efficient construction of reversible jump Markov chain Monte Carlo proposal distributions.
J. R. Stat. Soc., Ser. B, Stat. Methodol. 65(1), 3--39.
\vspace{0.1in}

\item[]  Chernozhukov, V. \& Hong, H. (2002), An MCMC approach to classical estimation, Journal of Econometrics, 114, 293-346.
\item[] Dempster, A.P. (1972). Covariance Selection. Biometrics 28: 157--175.
\vspace{0.1in}

\item[] Diaconis, P., \& Ylvisaker, D. (1979). Conjugate Priors for Exponential Families. Annals of Statistics, 7: 269--281.
\vspace{0.1in}

\item[] Dobra, A., Hans, C., Jones, B., Nevins, J. R., Yao, G., \& West, M. (2004). Sparse graphical models for exploring gene expression data. Journal of Multivariate Analysis, 90(1), 196-212.
\vspace{0.1in}

\item[] Finegold, M., \& Drton, M. (2011). Robust graphical modeling of gene networks using classical and alternative t-distributions. The Annals of Applied Statistics, 1057-1080.
\vspace{0.1in}

\item[]  Friedman, N. (2004). {Inferring cellular networks using probabilistic graphical models}. Science, 303, 799--805.
\vspace{0.1in}

\item []Friedman, J., Hastie, T.,  \& Tibshirani, R. (2008).  Sparse inverse covariance estimation with the graphical lasso. Biostatistics 9: 432--441.
\vspace{0.1in}

\item[] George, E. I., \& McCulloch, R. E. (1993). Variable selection via Gibbs sampling. Journal of the American Statistical Association, 88(423), 881--889.
\vspace{0.1in}

%
\item[] Ghosal, S., Ghosh, J. K., \& Van Der Vaart, A. W. (2000). Convergence rates of posterior distributions. Annals of Statistics, 28(2), 500--531.
\vspace{0.1in}

\item[] Giudici, P. (1996). Learning in graphical Gaussian models. Bayesian Statistics, 5, 621-628.
\vspace{0.1in}

\item[] Giudici, P., \& Green, A. P. (1999). Decomposable graphical Gaussian model determination. Biometrika, 86(4), 785-801.
\vspace{0.1in}

\item[] Hilger, R. A., Scheulen, M. E., \& Strumberg, D. (2002). The Ras-Raf-MEK-ERK pathway in the treatment of cancer. Oncology Research and Treatment, 25(6), 511-518.
\vspace{0.1in}

\item[] Jiang, W. (2005). Bayesian variable selection for high dimensional generalized linear models.
Technical Report 05-02, Dept. Statistics, Northwestern Univ.
\vspace{0.1in}

\item[] Jiang, W. (2007). Bayesian variable selection for high dimensional generalized linear models: convergence rates of the fitted densities. The Annals of Statistics, 35(4), 1487-1511.
\vspace{0.1in}

\item[] Kleijn, B. J., $\&$ van der Vaart, A. W. (2006). Misspecification in infinite-dimensional Bayesian statistics. The Annals of Statistics, 837-877.
\vspace{0.1in}

\item[]  Koenker, R. (2004). Quantile Regression for Longitudinal Data.  Journal of Multivariate Analysis, 91: 74--89.
\vspace{0.1in}

\item[] Koenker, R., \& Bassett Jr, G. (1978). Regression quantiles. Econometrica: journal of the Econometric Society, 33-50.
\vspace{0.1in}
\item[] Kotz, S., \& Nadarajah, S. (2004). Multivariate t-distributions and their applications. Cambridge University Press.
\vspace{0.1in}

\item[]  Kozumi, H., \& Kobayashi, G. (2011). Gibbs sampling methods for Bayesian quantile regression. Journal of statistical computation and simulation, 81(11), 1565-1578.
\vspace{0.1in}

\item[]   Kuo, L., \& Mallick, B. (1998). Variable selection for regression models. Sankhya Ser. B, 60: 65--81. 
\vspace{0.1in}

\item[] Lauritzen, S. L. (1996). Graphical Models. Claredon, Oxford. 
\vspace{0.1in}

\item[] {Li,H.,$\&$ Gui, J.}(2006). {Gradient directed regularization for sparse Gaussian concentration graphs with applications to inference of genetic networks}. Biostatistics. Vol 7, 302--317.
\vspace{0.1in}

\item[]  Li, Q., Xi, R., \& Lin, N. (2010).  Bayesian regularized quantile regression. Bayesian Analysis , 5 ,  3: 533--556. doi:10.1214/10-BA521. 
\item[] Liechty, J. C., Liechty, M. W., \&  M\"{u}ller, P. (2004). Bayesian correlation estimation. Biometrika, 91(1), 1-14.
\vspace{0.1in}

\item[] Liu, H., Han, F., Yuan, M., Lafferty, J.,\& Wasserman, L. (2012). High-dimensional semiparametric Gaussian copula graphical models. The Annals of Statistics, 40(4), 2293-2326.
\vspace{0.1in}

\item[] Mallick, B., Gold, D., \& Baladandayuthapani, V. (2009).  {{Bayesian} analysis of Gene expression data}. Wiley. 
\vspace{0.1in}

\item[] Meinshausen, N., \& B\"{u}hlmann, P. (2006). High-dimensional graphs and variable selection with the lasso. The Annals of Statistics, 1436--1462.
\item[] Neville, S.E., Ormerod, J.T., $\&$ Wand, M.P. (2014) Mean field variational Bayes for continuous sparse signal shrinkage: pitfalls and remedies. Electronic Journal of Statistics, 8, 1113-1151.
\vspace{0.1in}

\item[]  Peng, J., Wang, P., Zhou, N., \& Zhu, J. (2009). Partial correlation estimation by joint sparse regression models. Journal of the American Statistical Association, 104(486).
\vspace{0.1in}

\item[] Roverato, A. (2000). Cholesky decomposition of a hyper inverse Wishart matrix. Biometrika, 87(1), 99-112.
\vspace{0.1in}

\item[] Ryu, J. S., Memon, A., \& Lee, S. K. (2014). ERCC1 and personalized medicine in lung cancer. Annals of translational medicine, 2(4).

\vspace{0.1in}

\item[] Sch\"{a}fer, J., \& Strimmer, K. (2005). A shrinkage approach to large-scale covariance matrix estimation and implications for functional genomics. Statistical applications in genetics and molecular biology, 4(1).
\vspace{0.1in}

\item[] Scott, J. G., \& Carvalho, C. M. (2008). Feature-inclusion stochastic search for Gaussian graphical models. Journal of Computational and Graphical Statistics, 17(4).
\vspace{0.1in}

\item[] Scott, J. G., \& Berger, J. O. (2010). Bayes and empirical-Bayes multiplicity adjustment in the variable-selection problem. The Annals of Statistics, 38(5), 2587-2619.
\vspace{0.1in}

\item[] Segal, E., Shapira, M., Regev, A., Pe'er, D., Botstein, D., Koller, D., \& Friedman, N. (2003). Module networks: identifying regulatory modules and their condition-specific regulators from gene expression data. Nature genetics, 34(2), 166-176.
\vspace{0.1in}

%
\item[] Sriram, K., Ramamoorthi, R. V., \& Ghosh, P. (2013). Posterior consistency of Bayesian quantile regression based on the misspecified asymmetric Laplace density. Bayesian analysis, 8(2), 479--504.
\vspace{0.1in}

\item[] Wand, M.p., Ormerod, J.T., Padoan, S.A., $\&$ Fruhwirth, R. (2011). Mean field variational Bayes for elaborate distributions. Bayesian Analysis, 6, 847-900.
\vspace{0.1in}

%
\item[] Wang, H. (2012). Bayesian Graphical Lasso Models and Efficient Posterior Computation.  Bayesian Analysis, 7, 771--790.
\vspace{0.1in}

\item[] Wong, F., Carter, C. K., \&  Kohn, R. (2003). Efficient estimation of covariance selection models. Biometrika, 90(4), 809-830.
\vspace{0.1in}

\item[] Yang, E., Allen, G., Liu, Z., \& Ravikumar, P. K. (2012). Graphical models via generalized linear models. In Advances in Neural Information Processing Systems, 1358-1366.
\vspace{0.1in}

\item[]Yang, Y., Wang, H. \& He, X. (2015). Posterior inference in Bayesian quantile regression with asymmetric Laplace likelihood, International Statistical Review.
\vspace{0.1in}
\item[] Yatabe, Y., Takahashi, T., \& Mitsudomi, T. (2008). Epidermal growth factor receptor gene amplification is acquired in association with tumor progression of EGFR-mutated lung cancer. Cancer research, 68(7), 2106-2111.
\vspace{0.1in}

\item[] Yuan, M., \& Lin, Y. (2007). Model selection and estimation in the Gaussian graphical model. Biometrika, 94(1), 19-35.
\vspace{0.1in}

\item[]  Zhao, T., Liu, H., Roeder, K., Lafferty, J., \& Wasserman, L. (2012). The huge package for high-dimensional undirected graph estimation in R. The Journal of Machine Learning Research, 13(1), 1059--1062.

\end{enumerate}

\end{document}